# Mapping the Scholarship of Dark Pattern Regulation: A Systematic Review of Concepts, Regulatory Paradigms, and Solutions from an Interdisciplinary Perspective

Weiwei Yi[a], Zihao Li[b]

CREATe Centre, University of Glasgow

**Abstract**

Dark patterns—design tricks used on online interfaces to manipulate users' decision-making process—have raised public concerns. However, research on regulation of dark pattern remains underdeveloped and scattered, particularly regarding scholars' views on the concept, regulatory paradigms, and solutions. Following PRISMA guidelines, this paper systematically reviews the formats and content of regulatory discussions on dark patterns from the interdisciplinary scholarship of Law and Human-Computer Interaction. A total of 65 studies were analysed through content and thematic analysis. This study synthesises the unique trends and characteristics of legal scholarship on dark patterns, identifying five root problems and triple layered harms. It critiques current regulations in terms of legal theories and sectoral legislations, highlighting their inadequacies in addressing dark patterns. The paper also critically examines existing proposed solutions, including paradigmatic shifts in legal doctrines, refinements to existing frameworks, technical design-embedded solutions, and accountability measures for design practices. This research critically discusses the current barriers to effective dark pattern regulations and explores promising regulatory solutions. The difficulty in identifying the normative nature of various forms of dark patterns, in identifying evident and actionable harm, and the expanding scope of dark patterns' connotation inherently hinders effective regulation. However, technical design-embedded solutions, accountability frameworks, and practical design guidelines offer potential routes for more proactive regulation, while legal pluralism stands as a promising macro-level change in regulatory paradigms for dark pattern regulation.

[a] Weiwei Yi, PhD Candidate at CREATe Centre, School of Law, University of Glasgow, UK. We sincerely thank Dr. Amy Thomas and Professor Martin Kretschmer for their invaluable reviews and support throughout the writing of this paper. We are also deeply grateful to Leiden University's Center for Law and Digital Technologies for the opportunity to present at the "Law and/versus Technology: Trends for the New Decade" Conference in June 2024. We especially thank Dr. Arianna Rossi for her insightful comments that improved this paper.
[b] Zihao Li is a Lecturer in Law and Technology at CREATe Centre, School of Law, University of Glasgow, UK; He is also a Fellow at Stanford Law School, Stanford University, US. Email address: zihao.li@glasgow.ac.uk

# 1 Introduction

In the digital age, people interact daily with online services through their online interface and system design.[1] However, few scrutinise the deliberate design choices behind these interfaces.[2] This lack of awareness is exploited by UI designers to create "dark patterns" — 'tricks used in websites and apps that make you do things that you didn't mean to, like buying or signing up for something.'[3]—which causes various harms to consumer welfare, privacy, and beyond.[4] Including well-recognized specific dark pattern types (e.g., Roach Motel, bait and switch, Trick questions)[5], the overall dark pattern phenomenon is increasingly discussed against the bigger backdrop of a socio-technological approach to technology and under the issue of modifications to users' choice architecture in and beyond the context of user interfaces.[6]

More recently, dark patterns have increasingly become a subject of discussion among regulators and legal researchers.[7] Yet, despite the rapid growth of studies on dark patterns, particularly in the field of Human-Computer Interaction (HCI), researchers primarily focus is limited to identifying dark patterns across different contexts and refining taxonomies, paying less attention to normative evaluations and regulatory discussions.[8] This is because dark patterns affect a broad spectrum of interests and values protected by various laws, yet existing regulatory frameworks have proven inadequate in addressing them effectively. As will be demonstrate later through the review, one of the major challenges lies in the inherent ambiguity of the concept of "dark patterns" that complexify further regulatory discourse. Additionally, while privacy concerns are often at the forefront of regulatory efforts, other aspects such as consumer protection, competition are frequently receive less attention. Meanwhile, what consists of, and the alleged effectiveness, of the existing regulatory solutions remains clouded. In sum, research on how dark pattern regulation is portrayed by these researchers remains underdeveloped and warrants a comprehensive and interdisciplinary review. To address this gap and also bridge the knowledge differences on regulation between two crucial disciplines, this article systematically reviews existing academic literature on the regulation of dark patterns from both HCI and legal perspectives.[9]

---

[1] Gaia Bernstein, 'A Window of Opportunity to Regulate Addictive Technologies' (2022) 2022 Wisconsin Law Review Forward 64.
[2] Daniel Susser, 'Invisible Influence: Artificial Intelligence and the Ethics of Adaptive Choice Architectures', *Proceedings of the 2019 AAAI/ACM Conference on AI, Ethics, and Society* (ACM 2019) 404 <https://dl.acm.org/doi/10.1145/3306618.3314286> accessed 22 October 2023.
[3] As are various definitions of 'dark patterns', the paper ushers in the concept by referring to the preliminary definition provided by Harry Brignull, who first coined the term 'dark patterns'. See, Brignull, H, Leiser, M, Santos, C & Doshi, K, 'Deceptive Design' (2023) https://www.deceptive.design/ accessed 5 Jul 2024.
[4] Arunesh Mathur, Mihir Kshirsagar and Jonathan Mayer, 'What Makes a Dark Pattern... Dark? Design Attributes, Normative Considerations, and Measurement Methods', *Proceedings of the 2021 CHI Conference on Human Factors in Computing Systems* (Association for Computing Machinery 2021) <https://doi.org/10.1145/3411764.3445610> accessed 23 February 2023.
[5] Colin M Gray, Cristiana Santos and Nataliia Bielova, 'Towards a Preliminary Ontology of Dark Patterns Knowledge', *Extended Abstracts of the 2023 CHI Conference on Human Factors in Computing Systems* (ACM 2023) <https://dl.acm.org/doi/10.1145/3544549.3585676> accessed 9 June 2023.
[6] Kerstin Bongard-Blanchy and others, "I Am Definitely Manipulated, Even When I Am Aware of It. It's Ridiculous!" - Dark Patterns from the End-User Perspective', *Designing Interactive Systems Conference 2021* (ACM 2021) 763 <https://dl.acm.org/doi/10.1145/3461778.3462086> accessed 14 September 2023. See also Daniel Susser, Beate Roessler and Helen Nissenbaum, 'Technology, Autonomy, and Manipulation' (2019) 8 Internet Policy Review 7–8 <https://policyreview.info/articles/analysis/technology-autonomy-and-manipulation> accessed 4 October 2023.
[7] Gray, Santos and Bielova (n 5) 1.
[8] See e.g., Colin M Gray and others, 'Mapping the Landscape of Dark Patterns Scholarship: A Systematic Literature Review', *Companion Publication of the 2023 ACM Designing Interactive Systems Conference* (Association for Computing Machinery 2023) 192 <https://dl.acm.org/doi/10.1145/3563703.3596635> accessed 29 October 2023; Mathur, Kshirsagar and Mayer (n 4) 13.

## 1.1 Literature Review

Previous literature reviews on dark patterns have not provided a critical interdisciplinary systematic review focused on regulatory discussions. For example, reviews led by HCI researchers have conducted scoping reviews focusing on definitions, taxonomies, and normative theories, where discussions on regulations were limitedly touched upon,[10] or focused solely on empirical research,[11] or on research frameworks[12] where argumentative discussion on regulations were excluded. In existing organization-led reviews on dark patterns, either the relevance of results on dark patterns was jeopardized by the over-inclusion of other terms (e.g., Manipulative Persuasion) in the data search,[13] or the systematic review did not comprehensively address the regulations considering sectoral and paradigmatic differences between regulations.[14] Our systematic review contributes by thoroughly reviewing regulatory discussions on dark patterns within interdisciplinary scholarship to informing imminent policy.

## 1.2 Research Questions

This paper systematically reviews the formats and content of the regulatory discourse on dark patterns from the interdisciplinary perspectives of Law and HCI. This paper addresses four main questions:

- RQ1: What are the key characteristics of the scholarship on dark pattern regulation? (Section 3.1)
- RQ2: How does legal scholarship interact with and interpret the concept of 'dark patterns'?, (Section 3.2)
- RQ3: What specific regulatory frameworks, doctrines, and toolkits does the literature propose for regulating dark patterns? (Section 3.3)
- RQ4: How effective are existing and proposed regulations and toolkits in addressing the dark pattern issues? (Section 4)

## 1.3 Primary Contributions

Our primary contributions are as follows:

---

[9] We adopted a broader interpretation of 'regulation' inspired by the concept of 'lex informatica,' which views technical capabilities and system design choices as regulatory methods. See Joel R. Reidenberg, 'Lex Informatica: The Formulation of Information Policy Rules through Technology ' (1997-1998) 76 Tex L Rev 553. In the same vein, this systematic review includes HCI papers of regulation discussions to comprehensively reflect regulatory insights across both disciplines.
[10] Mathur, Kshirsagar and Mayer (n 4).
[11] Gray and others, 'Mapping the Landscape of Dark Patterns Scholarship' (n 8).
[12] Vibhav Singh, Niraj Kumar Vishvakarma and Vinod Kumar, 'Unveiling Digital Manipulation and Persuasion in E-Commerce: A Systematic Literature Review of Dark Patterns and Digital Nudging' [2024] Journal of Internet Commerce 1.
[13] Directorate-General for Justice and Consumers (European Commission) and others, *Behavioural Study on Unfair Commercial Practices in the Digital Environment: Dark Patterns and Manipulative Personalisation : Final Report* (Publications Office of the European Union 2022) <https://data.europa.eu/doi/10.2838/859030> accessed 29 October 2023.
[14] OECD, 'Dark Commercial Patterns' (OECD 2022) <https://www.oecd-ilibrary.org/science-and-technology/dark-commercial-patterns_44f5e846-en;jsessionid=WoBu8wGTVok0VBMLRk0DFIwvMSA_ogOBoZAFGcrQ.ip-10-240-5-52> accessed 25 July 2023.

- We systematically review and identify the main characteristics of the interdisciplinary scholarship on dark pattern regulation in law and HCI, concerning their frames and aims, contextual themes, jurisdictions.
- We demonstrate the unique trends and characteristics of legal scholarship approaching 'dark patterns' (e.g., regarding its positioning in the papers, its types and taxonomies, the scalability of its connotation) and illuminate the five root problems and triple layered harm of dark patterns as identified in legal studies.
- We synthesise the problems with current regulations (i.e., legal theories, departmental laws, governance beyond hard laws) in handling dark patterns and outline the primarily proposed regulatory solutions either paradigmatically, supplementally, or technically.
- We critically evaluate the current barriers to effective dark pattern regulations by interrogating the nature, harm and scope of the concept, explore potential solutions and some points for future research.

# 2 Methodology

## 2.1 Search Strategy

### 2.1.1 Collecting and Screening Relevant Literature

Considering that dark patterns are predominantly discussed in the research fields of information technology law and HCI, we utilize interdisciplinary databases that particularly relate to these two fields: JSTOR, HeinOnline, Scopus, Web of Science, and ACM Digital Library. To collect relevant literature, we employ search engines to retrieve only peer-reviewed, English-written journal articles, conference proceedings, and book chapters through an advanced search using combined keywords: "dark pattern*" OR "manipulative pattern*" OR "deceptive pattern*" OR "deceptive design*" OR "manipulative design*".[15] Wherever possible, the combined keywords are searched through "title, abstract, and indexing/keywords" (Web of Science and Scopus); where not available, "full text" is searched (ACM Digital Library, JSTOR, HeinOnline) to capture title and abstracts. Similarly, to further target the regulatory discourse of dark patterns, we specify the research fields of 'Law' and 'HCI' whenever possible. As illustrated in Figure 1, on 14th November 2023, we retrieved in 586 articles before the first screening after remove duplications.

## 2.2 Exclusion and inclusion criteria

---

[15] These selected keywords are commonly used substitutes for 'dark patterns' by scholars.

### 2.2.1 First round screening: inclusion criteria

In this step, we started the first-round screening manually with the following inclusion criteria, based on the article's title, abstract, and indexing keywords. For articles that we deem potentially relevant but with limited details, we tag them as "Maybe" and decide jointly after a process of full-text reading and discussion.[16]

1) The article should explicitly use the term "dark pattern(s)" or substitutes (i.e., deceptive/manipulative design(s)/pattern(s)).
2) The article should explicitly contain regulatory discussions regarding dark patterns (including legal and regulatory theories).

After the first screening, we excluded 476 articles that did not meet the inclusion criteria, leaving 110 papers for a second, full-text screening. Additionally, during this phase, we excluded non-papers, call-for-papers, workshops, papers not written in English, and those that were not accessible.

### 2.2.2 Second Round Screening: Inclusion Criteria and Division of Paper Types

We divide the literature into Group 1: law papers (including empirical and theoretical) and Group 2: HCI papers (empirical and theoretical). Papers are classified as either type according to the primary research topic they examine or evaluate.[17] When the distinction is not immediately apparent, we identify papers in which legal issues are discussed throughout the text as law papers to enrich our analysis.[18] To emphasize the focus on regulatory discourse on dark patterns, we deploy a second set of inclusion criteria for HCI papers:

1) The article must explicitly address policymakers or legal researchers. (i.e., we do not infer such intention from the content. Therefore, papers only focusing on 'design ethics' or 'better designs' or that only address HCI designers/communities would not fall within the scope).
2) The article substantially engages with legal or regulatory discussions (i.e., articles that merely 'call for more legal attention to the focused topic,' or simply promote detection methods and claim that 'this approach can inspire legal discussions/is helpful for policymakers,' or only mention legal aspects as descriptive backgrounds without substantial discussion of regulation will not be included).

---

[16] Different strategies are adopted for various databases to focus on papers containing regulatory discussions. For papers from HeinOnline, which are more law-oriented and often feature substantial regulatory discussions, we examine how keywords appear in these papers. For papers from the ACM Library that are more relevant to computer and information science, we particularly assess whether the words in the abstracts signal potential regulatory discussion and verify this through full-text reading. For the other three interdisciplinary databases where 'search by disciplines' is available, we lock on HCI/CS and Law/Social Science and similarly look for relevant keywords in abstracts..

[17] Our research includes both law and HCI for dark patterns' regulatory discussion rather than law papers alone. This is because, Subjectively, the review aims to bridge the dark pattern regulation knowledge that benefits both HCI and law sides and explore the potential for regulatory methods of technical characteristics; objectively, the HCI discipline remains the pioneer of dark pattern research and may pinpoint the regulatory limitations and directions in a grounded and practical manner, while the continuing collaboration within two disciplines blurs the differentiation and entails the consideration of papers based on content than origin.

[18] For instance, we define the following as law papers: Jamie Luguri and Lior Jacob Strahilevitz, 'Shining a Light on Dark Patterns' (2021) 13 Journal of Legal Analysis 43.and Colin M Gray and others, 'Dark Patterns and the Legal Requirements of Consent Banners: An Interaction Criticism Perspective', *Proceedings of the 2021 CHI Conference on Human Factors in Computing Systems* (ACM 2021) <https://dl.acm.org/doi/10.1145/3411764.3445779> accessed 9 December 2023. The following references define empirical research and empirical law papers: Malcolm Williams, *The SAGE Dictionary of Social Research Methods* (SAGE Publications, Ltd 2006) 91–92 <https://methods-sagepub-com.ezproxy1.lib.gla.ac.uk/reference/the-sage-dictionary-of-social-research-methods> accessed 23 May 2024.; Peter Cane and Herbert M Kritzer (eds), *The Oxford Handbook of Empirical Legal Research* (Oxford University Press 2010) 2 <https://doi.org/10.1093/oxfordhb/9780199542475.001.0001> accessed 23 May 2024.

After the second screening, 65 papers were retained, with 50 law papers (40 theoretical papers, 10 empirical papers) and 15 HCI papers (3 theoretical HCI papers, 12 empirical HCI papers).

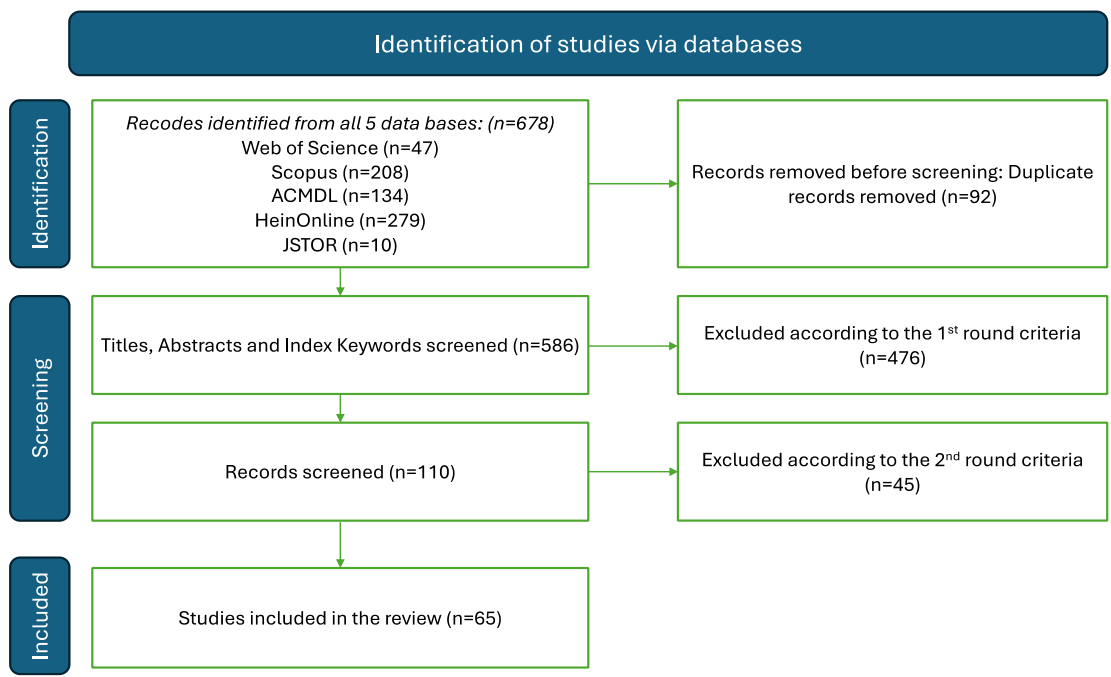

*Figure 1 PRISMA Diagram*

## 2.3   Literature Analysis

Given the diverse research questions and methodologies, we use both content and thematic analysis to address RQs with appropriate detail.[19] Content analysis is used to code directly identifiable and quantifiable answers to sub-questions in RQ1 (frames and aims, jurisdictions, and contextual themes), the first two sub-questions in RQ2 (role of dark patterns; engagement with definitions and taxonomies), and the first sub-question in RQ3 (main regulations discussed). We refer to Gray et al.'s codebook for HCI papers and answer the first sub-question in RQ1 deductively. [20] A pilot study over 35 random papers was conducted to generate the codebook inductively (See Annex). Codes for Q3 are non-exclusively distributed,[21] while other questions are exclusively coded based on relevance.

---

[19] Content analysis categorizes text items into codes, counting their occurrences and comparing them with other codes to draw observations from the document, with each code meticulously described and stored in a codebook for review, whereas thematic analysis is a classic qualitative research method to systematically identify, analyze, and interpret recurring patterns or themes within data, commonly focusing on subjective perspectives, where the data segments are organised by shared characteristics to construct cohesive themes that elucidate pertinent aspect of the dataset. The content analysis and thematic analysis used in this research are informed by the following references: Virginia Braun and Victoria Clarke, 'Can I Use TA? Should I Use TA? Should I Not Use TA? Comparing Reflexive Thematic Analysis and Other Pattern-Based Qualitative Analytic Approaches' (2021) 21 Counselling and Psychotherapy Research 37; Lisa Webley, *Qualitative Approaches to Empirical Legal Research* (Oxford University Press 2010) <https://academic.oup.com/edited-volume/35077/chapter/299091384> accessed 23 May 2024.

[20] The 'Framing' section of Gray et al.'s codebook is employed, with the title modified to 'frames and aims'. This section of the codebook is updated by the addition of a new code, 'theoretical', in order to reflect HCI papers that are non-empirical and are not included by the initial codes. This is done in order to make the codebook more suitable to the research in question. See the original codebook in Gray and others, 'Mapping the Landscape of Dark Patterns Scholarship' (n 8) 191.

Meanwhile, we use thematic analysis to code RQs whose answers are less quantifiable and emerge as layers of themes (beyond surface-level) that requires a substantial amount of interpretation. These include the last two sub-questions in RQ2 (root problems and harms of dark patterns) and the last two in RQ3 (identified regulatory problems and proposed solutions). We use open and axial coding to generate themes and answer each question.[22] After pilot coding, the two authors coded the remaining data using the established codebooks for content analysis and reconciled emerging themes through discussions for thematic analysis.

# 3 Findings

## 3.1 The basic characteristics of the literature (Frames and aims, Contextual themes, Jurisdictions)

For the 50 law papers, most aim to *assess the effectiveness of existing legislation* (38%), while *evaluations of the legitimacy* of a practice are least framed (14%). Comparatively, among the 15 HCI papers, the majority are likely to be *taxonomy-building* (33.34%) for instance, synthesizing a collective multi-layer taxonomy of dark patterns using regulatory and academic literature[23], or identifying new types of dark patterns in multi-modality or live stream e-commerce.[24] Papers with *descriptive* framing or *detection-focused* are fewer (each at 13.3%), while only 6.6% are *experimental/causal*, and none are *evaluative*. Additionally, we identify *theoretical* papers (20%) that do not contain empirical elements.[25] Therefore, similar to Gray et al.'s review of HCI empirical scholarship, our findings also reflect the diversity of framing contributions within dark pattern literature, although we observe more taxonomy-building framing and no evaluative framings in our data. However, considering that regulatory discussions are likely to be relatively independent additions in the discussion sections, it remains to be confirmed whether there is any pattern of relevance between the substantiality of regulatory discussions and the frames and aims of a study. As signalled earlier, we also observed the difficulty to classify some papers as purely 'HCI' or 'law', which again reflects interdisciplinary nature of 'dark patterns' that requires both the normative assessment of 'dark(ness)' and empirical identification and measurements of design 'patterns', which leads to more and deep collaborations from both disciplinary specialities, and is accompanied by the ongoing movement of empirical studies in legal scholarship.

---

[21] This means that more than one code can be distributed to a paper to answer the same question.
[22] Axial coding is a method used to establish connections between categories and subcategories within qualitative data. It forms linkages that unveil emergent themes and facilitate the creation of new categories or subcategories. This process enables researchers to make theoretical claims about the patterns and relationships found in their data. See Nathaniel Simmons, 'The SAGE Encyclopedia of Communication Research Methods' in pages 80-82 (SAGE Publications, Inc 2024) <https://methods.sagepub.com/reference/the-sage-encyclopedia-of-communication-research-methods>.
[23] Gray, Santos and Bielova (n 5).
[24] J Gunawan and others, 'A Comparative Study of Dark Patterns across Web and Mobile Modalities' (2021) 5 Proceedings of the ACM on Human-Computer Interaction <https://www.scopus.com/inward/record.uri?eid=2-s2.0-85117880047&doi=10.1145%2f3479521&partnerID=40&md5=e430f5a284b909046021abd5dea558cf>; Qunfang Wu and others, 'Malicious Selling Strategies in Livestream E-Commerce: A Case Study of Alibaba's Taobao and ByteDance's TikTok' (2023) 30 ACM TRANSACTIONS ON COMPUTER-HUMAN INTERACTION.
[25] e.g., Arvind Narayanan and others, 'Dark Patterns: Past, Present, and Future: The Evolution of Tricky User Interfaces' (2020) 18 Queue Pages 10:67. Mathur, Kshirsagar and Mayer (n 4); D Susser and V Grimaldi, 'Measuring Automated Influence: Between Empirical Evidence and Ethical Values' (2021) <https://www.scopus.com/inward/record.uri?eid=2-s2.0-85112465535&doi=10.1145%2f3461702.3462532&partnerID=40&md5=9b2362c93700efc8b8eaefe271c86d1d>.

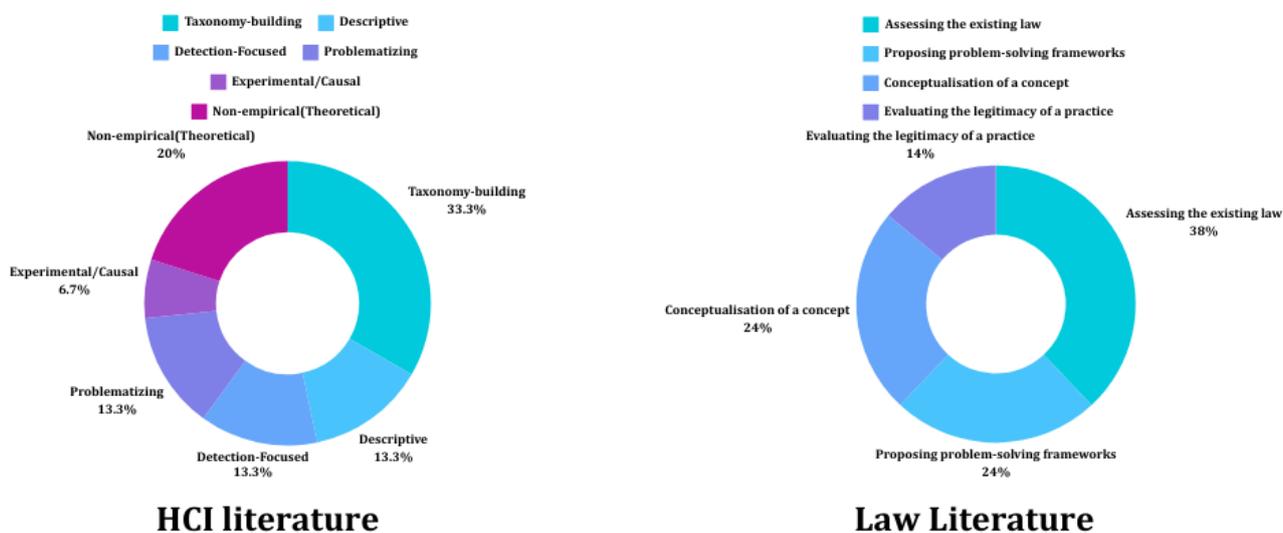

*Figure 2 Frames and Aims*

Regarding jurisdictions, apart from studies that show *multi-jurisdictional characteristics* (32.3%), most studies focus on the *United States* (33.8%), while the *European Union* (either at the EU level or national level) represents the third most studied region (27.7%). A minority of studies cover *the United Kingdom* (3.1%), *Canada* (1.5%), and *China* (1.5%). Clearly, although many papers do not specify jurisdictions, the United States and the European Union dominate the regulatory discourse on 'dark patterns' and related terms, influenced by the language of publication.

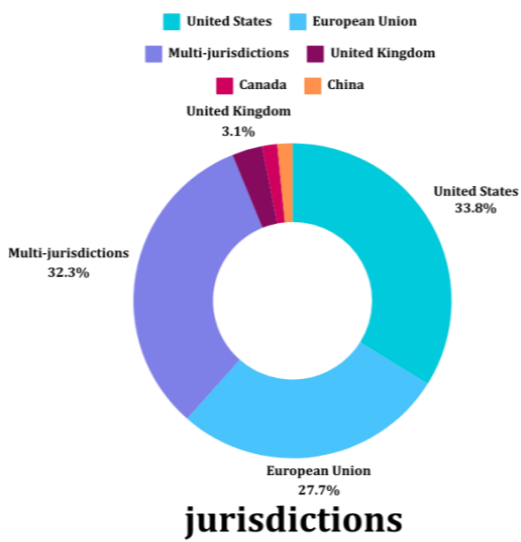

*Figure 3 Jurisdictions*

Regarding the contextual themes, the majority investigate dark patterns holistically and at a macro level (*Digital Services*) and tend to merely exemplify the diversity of dark patterns (38.5%).[26] Nonetheless, many articles still focus on micro-interactions with *user interfaces on websites or apps* (e.g., cookie banners/consent management interfaces, checkout processes on shopping websites, content reporting interfaces on apps) (30.8%).[27] Meanwhile, papers that emphasize the roles and *effects of platforms* or related subjects in dark pattern practices tend to be legal or theoretical (16.9%).[28] *Algorithms and artificial intelligence* (9.2%) and *games and gaming* (4.6%) remain the least discussed, but still stand out as emerging themes where regulatory discourse is seeing growth.[29] The varied focuses of papers, ranging from macro-level discussions on general impacts to micro-level analyses of user interactions within specific scenarios, reflect the multifaceted nature of dark pattern research across disciplines.

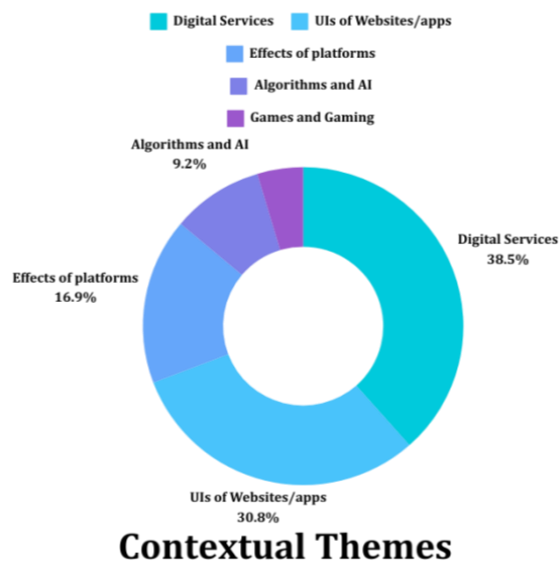

*Figure 4 Contextual Themes*

## 3.2   How does the law literature engage with dark pattern?

### 3.2.1   How is 'dark patterns' engaged with in the articles?

Among the 50 legal papers reviewed, those treating dark patterns *as the main focus* occupy the largest share (38%). Many papers position dark patterns *as a case or theme* discussed independently or alongside other major subjects (e.g., as one aspect of the usability of cookie consent interfaces,[30] or compared with behavioural advertising, personalisation, and recommendation systems[31])(30%). Other papers use dark patterns *as examples* (24%), included under broader subjects (e.g., 'platform nudges',[32] the concept of online manipulation[33]), which the papers primarily address. Some papers set dark patterns *as the context* and backdrop to introduce the main topics (8%).[34]

---

[26] See e.g., Kirsten Martin, 'Manipulation, Privacy, and Choice' (2021) 23 North Carolina Journal of Law & Technology 452; Lindsay Wilson, 'Is There a Light at The End of the Dark-Pattern Tunnel?' (2023) 91 GEORGE WASHINGTON LAW REVIEW 1048.
[27] See e.g., Hana Habib and others, '"Okay, Whatever": An Evaluation of Cookie Consent Interfaces' (Association for Computing Machinery 2022) <https://doi.org/10.1145/3491102.3501985>; Arunesh Mathur and others, 'Dark Patterns at Scale: Findings from a Crawl of 11K Shopping Websites' (2019) 3 Proc. ACM Hum.-Comput. Interact. <https://doi.org/10.1145/3359183> accessed 1 January 11AD; Ben Wagner and others, 'Regulating Transparency? Facebook, Twitter and the German Network Enforcement Act' (Association for Computing Machinery 2020) <https://doi.org/10.1145/3351095.3372856>.
[28] See e.g., Mark MacCarthy, 'Enhanced Privacy Duties for Dominant Technology Companies' (2021) 47 Rutgers Computer and Technology Law Journal 1.
[29] Tegan Cohen, 'Regulating Manipulative Artificial Intelligence' (2023) 20 Scripted; Sheldon A. Evans, 'Pandora's Loot Box' (2022) 90 The George Washington Law Review 376.

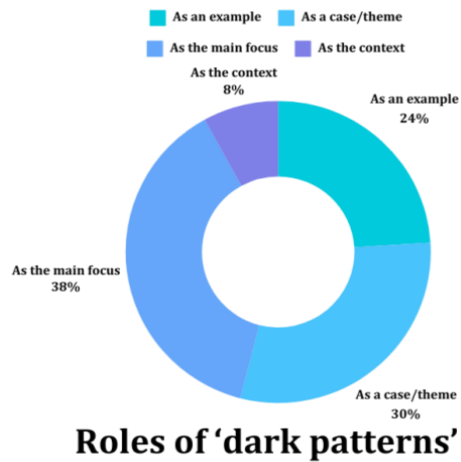

*Figure 5 Types of engagements with 'Dark patterns'*

### 3.2.2 Definitions and taxonomies of dark patterns

#### 3.2.2.1 References of definitions, approaches to dark pattern types and taxonomies

Legal scholarship on dark patterns predominantly relies on definitions from HCI scholarship rather than legal or policy documents. Brignull's definition is the most frequently referenced,[35] followed by Mathur et al.,[36] then Gray et al.,[37] and lastly, Luguri and Strahilevitz.[38] However, unlike the focus on taxonomies typical in HCI literature on dark patterns[39], we find that half of the papers (50%) *neither engage with the taxonomies* nor explicitly refer to specific HCI jargon or subcategories. Some papers (36%) only reference dark pattern types as examples, particularly when examining the effects of platforms, while only a few (14%) focus on taxonomy discussions as a research objective, whether proposing, refining,[40] or evaluating taxonomies.[41] This finding suggests that legal scholarship tends to examine and evaluate dark patterns as a holistic phenomenon and may not sufficiently discuss the diverse natures of different dark pattern types.

---

[30] K Lapin and L Volungevičiūtė, 'Improving the Usability of Requests for Consent to Use Cookies' (2023) <https://www.scopus.com/inward/record.uri?eid=2-s2.0-85172019400&doi=10.1007%2f978-3-031-37649-8_19&partnerID=40&md5=3559e324d63afc5f3c4a17324a442301>.
[31] Sébastien Fassiaux, 'Preserving Consumer Autonomy through European Union Regulation of Artificial Intelligence: A Long-Term Approach' [2023] European Journal of Risk Regulation 1.
[32] Abbey Stemler, Joshua E Perry and Todd Haugh, 'The Code of the Platform' (2019) 54 Georgia Law Review 605.
[33] Shaun B Spencer, 'The Problem of Online Manipulation' (2020) 2020 University of Illinois Law Review 959.
[34] See e.g., Sheldon A. Evans (n 29).
[35] Papers commonly refer to two sources. The first one is Harry Brignull, Dark Patterns: Dirty Tricks Designers Use to Make People Do Stuff, 90 PERCENT OF EVERYTHING (July 8, 2010), https://www.90percentofeverything.com/2010/07/08/dark-patterns-dirty-tricks-designersuse-to-make-people-do-stuff/ [https://perma.cc/3NES-WHWP]. The other one is from 'Deceptive Patterns' <https://www.deceptive.design/> accessed 8 March 2024.
[36] Most of the references are made to a study in 2019 (Mathur and others (n 27).) than the review in 2021 (Arunesh Mathur, Mihir Kshirsagar and Jonathan Mayer, 'What Makes a Dark Pattern... Dark? Design Attributes, Normative Considerations, and Measurement Methods' (Association for Computing Machinery 2021) <https://doi.org/10.1145/3411764.3445610>.)
[37] Colin M Gray and others, 'The Dark (Patterns) Side of UX Design', *Proceedings of the 2018 CHI Conference on Human Factors in Computing Systems* (ACM 2018) <https://dl.acm.org/doi/10.1145/3173574.3174108> accessed 19 January 2024.
[38] Jamie Luguri and Lior Jacob Strahilevitz, 'Shining a Light on Dark Patterns' 13 43.
[39] Gray and others, 'Mapping the Landscape of Dark Patterns Scholarship' (n 8) 190–191.

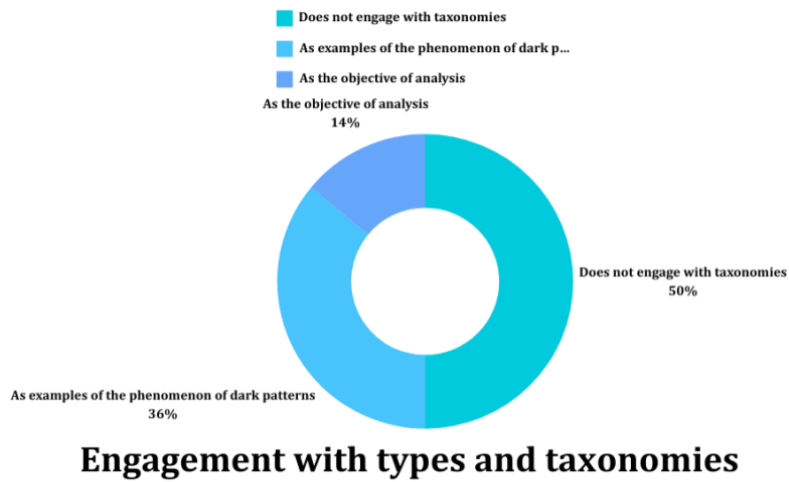

*Figure 6 Types and taxonomies of dark patterns*

### 3.2.2.2   The divergent scopes of dark patterns by some legal scholars

We have also noted discussions among legal researchers about expanding the concept of dark patterns to systemic issues, including personalized dark patterns.[42] For example, dark design patterns are being examined in non-traditional interfaces such as voice and gesture-based systems and virtual reality. These discussions explore how customization and user profiling could enhance the influence of dark patterns by targeting specific vulnerabilities of individuals or groups.[43] Additionally, elements like loot boxes in video games are identified as dark patterns that could be intensified through gamer profiling.[44] The intertwining of dark patterns with AI is noteworthy, with some researchers discussing 'second-generation dark patterns' employed by AI to create 'hypernudges,' or referring to the 'unacceptable risk systems' under the proposed EU AI Act as 'dark pattern systems'.[45] This broader understanding of dark patterns aligns with HCI researchers' views on advanced personalised UIs as the future of dark patterns, moving beyond the basic forms currently prevalent and exploited by companies.[46]

---

[40] See e.g., SD Conca, 'The Present Looks Nothing like the Jetsons: Deceptive Design in Virtual Assistants and the Protection of the Rights of Users' (2023) 51 Computer Law and Security Review <https://www.scopus.com/inward/record.uri?eid=2-s2.0-85171189474&doi=10.1016%2fj.clsr.2023.105866&partnerID=40&md5=db981cf8333de5023b6aecd8dbd1bedb>.
[41] See e.g., A Hung, 'KEEPING CONSUMERS IN THE DARK: ADDRESSING "NAGGING" CONCERNS AND INJURY' (2021) 121 Columbia Law Review 2483.
[42] The impact of the evolving diversity of dark pattern scope will further be evaluated in the discussion section.
[43] Mason Marks, 'Biosupremacy: Big Data, Antitrust, and Monopolistic Power over Human Behavior' (2021) 55 513, 535; Conca (n 40).
[44] Sheldon A. Evans (n 29) 428.
[45] Stefano Faraoni, 'Persuasive Technology and Computational Manipulation: Hypernudging out of Mental Self-Determination' (2023) 6 FRONTIERS IN ARTIFICIAL INTELLIGENCE 4; Vera Lúcia Raposo, 'Ex Machina: Preliminary Critical Assessment of the European Draft Act on Artificial Intelligence' 30 88.
[46] Narayanan and others (n 25) 79; Susser and Grimaldi (n 25).

### 3.2.3 What are the root problems/causes of dark patterns?

Through thematic analysis, we identify five root problems in this section that either derive from or catalyse the dark pattern problem: invisibility of digital design, platform effect and digital market competition, human bias and vulnerability, (attention-seeking) business models & interest conflict, and power imbalance over information and design.

*Invisibility of digital design*

It is commonly acknowledged that the online environment is an indispensable contextual component of dark pattern practices. The uniqueness of dark patterns lies in the infinite and instant malleability of software interfaces on online platforms, as design choices can be tweaked and updated through low-cost online experiments and automation tools. Additionally, the vast scale of potential gains from the direct result of expanding digitization and globalization stokes the prevalence and influence of dark patterns.[47] Besides commonly mentioned A/B testing, scholars also believe that technologies such as behavioural analytics, user profiling and targeting, and even artificial intelligence have gradually endowed more power to dark patterns over users.[48] As information technology "sits almost entirely outside conscious awareness," users seldom ponder the inner workings but still actively use it daily, dark patterns benefit from the same invisibility.[49] Even though the screens are tangible, designs that incentivise overuse and impulsive use remain invisible.[50] Such invisibility of dark patterns has profound implications for long-term thinking,[51] which may "increase the risk that society will not engage in interpretive flexibility" and enables the technology to rapidly reach the phase of closure, at which point few social reflections on the choices of design and use would occur."[52]

*Platform effect and digital market competition*

Many legal researchers approach the dark pattern phenomenon by examining the roles and effects of platforms in the digital market and tend to view dark patterns as an abuse of platform (market) power.[53] Compared to their brick-and-mortar counterparts, digital platforms are different as they benefit from fewer reputational concerns and competitive pressures since users commonly face inferior exit options and greater lock-in due to their investment in these platforms, and from little regulation regarding their harmful practices on consumers.[54]

---

[47] Justin Hurwitz, 'Designing a Pattern, Darkly' (2020) 22 North Carolina Journal of Law & Technology i, 69; L Barros and others, 'The Rise of Dark Patterns: Does Competition Law Make It Any Brighter?' (2022) 21 Competition Law Journal 136, 140; Hung (n 41) 4.
[48] Jon M Garon, 'Protecting Public Health Amidst Data Theft, Sludge, and Dark Patterns: Overcoming the Constitutional Barriers to Health Information Regulations' (2023) 56 Akron Law Review 206; Sheldon A. Evans (n 29) 428; Conca (n 40) 18; Faraoni (n 45) 4.
[49] Daniel Susser, Beate Roessler and Helen Nissenbaum, 'Online Manipulation: Hidden Influences in a Digital World' (23 December 2018) 33–34 <https://papers.ssrn.com/abstract=3306006> accessed 19 September 2023.
[50] Bernstein (n 1) 73.
[51] Fassiaux (n 31) 8.
[52] Bernstein (n 1) 73.
[53] Olivier Sylvain, 'A New Telecommunications Act: Prioritizing Consumer Protection and Equality Telecommunications Act of 1996 Symposium' (2022) 37 Berkeley Technology Law Journal 281, 305; Gregory Day and Abbey Stemler, 'Are Dark Patterns Anticompetitive?' (2020) 72 Alabama Law Review 1, 24.
[54] CN Griffin, 'SYSTEMICALLY IMPORTANT PLATFORMS' (2022) 107 Cornell Law Review 445, 475.

However, regarding how the market competition structure influences dark pattern practices, scholars' opinions diverge. Some argue that the increasing centralization of market power by a handful of platforms and the lack of competition invalidate the market's self-correction process, leading to a deficiency of viable alternatives for consumers. In this context, firms can design interfaces to addict, subtly influence, or manipulate users and 'force purchasers to do something they would not do in a competitive market'.[55] However, other academics argue that it is the ever-fiercer competition in the digital market that pushes companies to conduct exploitative design intentionally, arguing that in the presence of consumer cognitive bias and increased competition, companies could be passively forced to 'phish for phools' (e.g., using dark patterns to exploit consumers), 'because if they do not, they will be replaced by competitors that will'.[56]

*Human bias and vulnerability*

Bounded rationality, which links to more normative concepts like 'human bias' and 'vulnerabilities', is commonly recognized as the fundamental concept to understand the effectiveness of dark patterns. The theories sometimes referred to by legal researchers are the dual-system theory by Daniel Kahneman and Amos Tversky, or the nudge theory (including discussions of 'choice architecture', 'liberal paternalism', and 'nudge and sludge') by Thaler and Sunstein.[57] Scholars typically explain how dark patterns are designed to make use of various heuristics in conditions of imperfect information, to exploit individuals' limited time, attention, and brainpower, to push them to adopt rules of thumb for efficiency and quick solutions, and to 'program them for specific behaviour' at the cost of their long-term best interest.[58]

*(Attention-seeking) Business models and Interest conflict*

Tech companies' attention-seeking business models exploit human vulnerabilities in the digital economy, where attention is the primary currency. Platforms often offer low-cost or free services to attract attention, designing interfaces that nudge or hook users, creating addiction through dopamine rushes. This increases user engagement on the platform, facilitating greater personal data collection and ad sales.[59]

Meanwhile, the interest conflict between users and service providers is increasingly recognised as the main reason for the emergence and expansion of dark patterns.[60] Dark patterns arise as digital platforms do not hold the best interests of users in mind and show indifference, if not outright exploitation, towards users' privacy and other interests.⁶¹ When the 'growth hacking mantra' is adopted for a business to pursue solely the maximization of certain metrics like growth and engagement, dark patterns become inevitable and even welcomed among designers, as success is measured solely by business metrics related to ads, website, or app performance.⁶² Even in the absence of this aggressive mantra, the business model still poses a core dilemma for website owners to balance revenue needs and legal compliance.⁶³ This close-knit relationship between the freemium (low-price) business model and the derivative interest conflict undermines the classic corporate model between consumers and service/product providers, which is based on attracting consumers with desirable products at competitive prices.⁶⁴

---

[55] Peter O'Loughlin, 'Cognitive Foreclosure Rethinking Antitrust' (2021) 38 Georgia State University Law Review 1097, 1167; See also MacCarthy (n 28) 71–72; Hung (n 41).
[56] Lauren E. Willis, 'Deception by Design' (2020) 34 Harvard Journal of Law & Technology 115, 151; Barros and others (n 47) 141; Griffin (n 54) 475.
[57] See e.g., Day and Stemler (n 53); Stemler, Perry and Haugh (n 32); Luguri and Strahilevitz (n 18).
[58] Barros and others (n 47) 138–139; Kristina Lapin and Laima Volungevičiūtė, 'Improving the Usability of Requests for Consent to Use Cookies' in Cezary Biele and others (eds), *Digital Interaction and Machine Intelligence*, vol 710 (Springer Nature Switzerland 2023) 193–194 <https://link.springer.com/10.1007/978-3-031-37649-8_19> accessed 9 December 2023; Hurwitz (n 47) 62.
[59] Day and Stemler (n 53) 7–8; J Gunawan, C Santos and I Kamara, 'Redress for Dark Patterns Privacy Harms? A Case Study on Consent Interactions' (2022) 181 <https://www.scopus.com/inward/record.uri?eid=2-s2.0-85142525860&doi=10.1145%2f3511265.3550448&partnerID=40&md5=8c5d9c396f4943033a64f723b72a40cb>; Bernstein (n 1) 69–70.
[60] Ryan Calo, 'Digital Market Manipulation' (2013) 82 George Washington Law Review 995, 1044; Susser, Roessler and Nissenbaum, 'Online Manipulation' (n 49) 35; Bernstein (n 1) 74.

*Derivative problems: Power asymmetry over information and design*

The <u>imbalanced power of information</u> underpins the potency of dark patterns. Consumer data collected by platforms is the main source of the growing knowledge on individuals and groups that exacerbates the information imbalance and transforms into powers of manipulation. People may remain unaware that they have been subjects of A/B testing and online experiments, and that this information is used to create and refine dark patterns.⁶⁵ Ultimately, dark patterns utilising this information asymmetry enable vendors to have more or better information than users, creating other power imbalances that exploit consumers, such as in commercial transactions.⁶⁶

Comparably, the <u>asymmetry of design power</u> is another assistor of dark patterns. Even though it is gradually understood that 'Design is everything…[D]esign is power',⁶⁷ it is usually the platforms that hold the design power and act as choice architects, intentionally configuring how choices are presented and thus imposing significant impacts on user decision-making.⁶⁸ Platforms are given 'too much power', regarding the implementation, and this power imbalance results in practices ranging from nudging users to overlook cookie policies to implanting more complex dark patterns.⁶⁹ Such design power imbalance is continuously exacerbated as power continues being concentrated in a handful of technology companies, leading to weakened user participation and negotiation over feasible design alternatives.⁷⁰

Consequently, these observed factors collectively result in the emergence and prevalence of dark patterns in the digital environment. Contrary to minor opinions that see dark patterns as merely an online replication of offline unfair practices,[71] our findings stress the unprecedented feature of dark patterns supported by the digital environment and technologies, influenced by the unique landscape of digital market competition and the inherent dilemma of business models. Moreover, the power imbalance of information and design might have changed 'the face of market manipulation to the point that it strains consumer protection law',[72] which further leads to calls for regulatory paradigm changes to focus on dealing with the power imbalance (See Section 3.3.3).[73]

---

⁶¹ Woodrow Hartzog and Neil Richards, 'THE SURPRISING VIRTUES OF DATA LOYALTY' (2022) 71 EMORY LAW JOURNAL 36–37.
⁶² Lauren E. Willis, 'Deception by Design' (2020) 34 Harvard Journal of Law & Technology 115, 150.
⁶³ Gray and others, 'Dark Patterns and the Legal Requirements of Consent Banners' (n 18) 11.
⁶⁴ Caleb N Griffin, 'Systemically Important Platforms' 107 445, 30.
⁶⁵ Luguri and Strahilevitz (n 38) 45; Gray and others, 'Dark Patterns and the Legal Requirements of Consent Banners' (n 18) 12; Thomas D Haley, 'Illusory Privacy' (2022) 98 Indiana Law Journal 75, 95.
⁶⁶ Mark Leiser, 'Illuminating Manipulative Design: From "Dark Patterns" to Information Asymmetry and the Repression of Free Choice under the Unfair Commercial Practices Directive' 34 484, 506.
⁶⁷ Hung (n 41) 2486.
⁶⁸ Stemler, Perry and Haugh (n 32) 625.
⁶⁹ Gerhard Seuchter and others, 'THE CRUX OF COOKIES CONSENT: A LEGAL AND TECHNICAL ANALYSIS OF SHORTCOMMINGS OF COOKIE POLICIES IN THE AGE OF THE GDPR' (2020) 9.
⁷⁰ Konrad Kollnig and others, '"We Are Adults and Deserve Control of Our Phones": Examining the Risks and Opportunities of a Right to Repair for Mobile Apps' (Association for Computing Machinery 2023) 22 <https://doi.org/10.1145/3593013.3593973>.

### 3.2.4 What are the identified harms of dark patterns?

This section categorises the harms of dark patterns into three areas: individual, market and competition, and societal impacts.

#### 3.2.4.1 Harms to individuals

The most pronounced harms of dark patterns are to individuals. The <u>harm to individual decision-making ability</u> by imposing labour or cognitive burdens is extensively recognized, typically taking the form of constant distractions of attention, lengthy times required to complete tasks, and emotional distress, among others.[74] Comparatively, <u>monetary harm</u> is less emphasised in legal scholarship than in HCI scholarship. One specific harm related to the long-term effects of dark design patterns is <u>addiction</u>. It is claimed that companies have implemented dark design patterns (e.g., gamification, reinforcement reward design) that cause users to become addicted to the product, leading to various mental health issues. Researchers particularly stress the amplification of damage to vulnerable groups such as children.[75] Additionally, the recognition of harms shows regional disparity: addiction as a harm of dark patterns tends to be recognized in the US, while such harm is not well-recognized in the EU.

Legal scholars tend to associate dark patterns with harms to human rights and common values, such as <u>autonomy</u>, <u>privacy</u>, and <u>fairness</u>. In line with Gunawan et al.'s findings, almost all papers discuss, to some extent, the harm to individual autonomy or self-determination by impairing individuals' decision-making capacity.[76] Harm to privacy ranges from well-known cases of consent interface designs to the nascent case of private sphere surveillance.[77] Harms to the value of fairness between traders and consumers are also commonly discussed, alongside evaluations of consumer protection frameworks.[78]

---

[71] Gregory M Dickinson, 'Privately Policing Dark Patterns' 57 1633, 1648.
[72] Calo (n 60) 1049.
[73] Hartzog and Richards (n 61) 988.
[74] Gunawan, Santos and Kamara (n 59) 183; Susser, Roessler and Nissenbaum, 'Online Manipulation' (n 49) 64; Faraoni (n 45) 5; Day and Stemler (n 53) 17.
[75] Griffin (n 54); Kollnig and others (n 70); Peter O'Loughlin, 'COGNITIVE FORECLOSURE' (2022) 38 GEORGIA STATE UNIVERSITY LAW REVIEW; Mark Warner, 'The Need for Legal Rules on Dark Pattern Website Activities' (2019) 3 International Journal for the Data Protection Officer and Privacy Counsel 7; Leon Y Xiao, 'Shopping Around for Loot Box Presence Warning Labels: Unsatisfactory Compliance on Epic, Nintendo, Sony, and Microsoft Platforms' (2023) 1 Games: Research and Practice 1; Bernstein (n 1).
[76] Gunawan, Santos and Kamara (n 59) 183.
[77] Conca (n 40).
[78] See e.g., Lauren E. Willis (n 56); Leiser, M.R. and Caruana, M., 'Dark Patterns: Light to Be Found in Europe's Consumer Protection Regime' (2021) 10 Journal Of European Consumer And Market Law 237; Gunawan and others (n 24); Dickinson (n 71).

### 3.2.4.2 Harms to the market and competition

Discussion of harms from market and competition perspectives has grown since Mathur et al.'s review.[79] Some researchers argue that the harm to market competition stems from dark patterns interfering with market participants' (consumers') decision-making processes in transactions, which subsequently results in reduced market efficiency that accumulates into anti-competitive effects and harms consumer welfare and market trust, especially when dark patterns are deployed by players with market power.[80]

### 3.2.4.3 Harm to society

Dark patterns' harm to society primarily originates from their disruption to the effective functioning of laws. For instance, in the context of privacy and data protection, dark patterns compromise the consent mechanism, which is central to contemporary privacy and data protection regimes.[81] It is important to note that dark patterns can also subvert other regulatory objectives beyond the commonly acknowledged protection of consumer welfare or privacy. For example, dark design patterns used by Facebook in the platform's reporting procedures for hate speech and terrorist content have imposed a series of redirections and obstructions on reporters, degrading the implementation of Germany's Network Enforcement Act,[82] which may jeopardise democracy and national security. From a broader perspective, it is believed that 'dark patterns distract from two of the central objectives of the European Union's digital single market: creating the right conditions and a level playing field for digital networks and innovative services to flourish and maximise the growth potential of the digital economy'.[83] Additionally, some researchers believe that dark patterns undermine individuals' capacity for self-government and their ability to pursue their own goals, further harming democracy at a fundamental level.[84], Others discuss how dark design patterns in games (e.g., 'loot boxes') resemble unrestricted gambling in reality,[85] and how these design mechanisms could lead to similar negative externalities in society.[86]

## 3.3 How do the papers engage with regulatory discourse of dark patterns?

In this section, we review all papers to provide a concise overview of the legal and regulatory discourse on dark patterns, beginning with the regulations involved, identifying related problems, and concluding with proposed solutions.

---

[79] Mathur, Kshirsagar and Mayer (n 4) 11. This discrepancy in number might be because we also review competition law papers that put dark patterns as a theme or examples of a bigger topic, which substantially enriches the dark pattern scholarship by capturing relevant contextual and peripheral debates.
[80] See e.g., Dickinson (n 71) 1647–1648; Leiser, M.R. and Caruana, M. (n 78) 6; Hung (n 41) 2498–2501. According to our findings, to what extent the market power of platforms (designers) influences the deployment of dark patterns and the market competition structure is not settled in the legal literature. We discuss this in Section 4.1.2
[81] Jeremy Wiener, 'Deceptive Design and Ongoing Consent in Privacy Law' (2021) 53 133; Gray and others, 'Dark Patterns and the Legal Requirements of Consent Banners' (n 18); Lapin and Volungevičiūtė (n 58); Haley (n 65).
[82] Wagner and others (n 27).
[83] Leiser, M.R. and Caruana, M. (n 78) 6.
[84] Marks (n 43); Susser, Roessler and Nissenbaum, 'Online Manipulation' (n 49).
[85] Xiao (n 75).
[86] Sheldon A. Evans (n 29).

### 3.3.1 The regulations regarding dark patterns mainly discussed[87]

We identify seven clusters of laws, among them *data protection and privacy law* (n=27)[88] and *consumer protection law* (n=27)[89] are widely recognized as crucial for regulating dark patterns, with *Competition law* next (n=12).[90] *Contract law* and *fundamental laws* being less touched upon (each n=5). Discussions on contract law mainly focus on the theory of contracting that questions the nature of consent to privacy policies concerning dark patterns[91] or proposing a private law solution to dark patterns from the perspective of contract law.[92] Meanwhile, fundamental rights are sometimes mentioned either to show how human rights (e.g., the right to data protection) are infringed by dark patterns,[93] or how the regulation of dark patterns might face obstructions from the US First Amendment concerning commercial speech.[94]

There is a small number of *other laws* seen as viable regulations as potential tools for reigning in dark patterns (n=12), such as gambling law, intellectual property law, tort law including product liability law, financial law, health insurance law, and corporate law. Moreover, many papers express expectations for the *EU digital strategy package* (n=7), particularly the Digital Services Act (DSA), Digital Markets Act (DMA), Data Act, and Artificial Intelligence Act (AI Act), regarding more effective regulation of dark patterns and other computational manipulative practices.

However, even though there is a comparable number of research from a consumer protection perspective, empirical studies on dark patterns from a consumer welfare protection perspective are less researched,[95] and there is no empirical research from a competition law perspective. Therefore, further research on dark patterns could benefit from more empirical studies about the effects on platform markets and consumer welfare.[96]

---

[87] Codes are distributed non-exclusively in general, except for papers that only engage with theories without identifiable regulations. Correspondingly, the presentation of content analysis is in number than percentage.
[88] For example, General data protection regulation (including the previous Directive) and the e-Privacy directive (including the proposed regulation), the Children's Online Privacy Protection Act, the California Consumer Privacy Act, the California Privacy Rights Act, the Colorado Consumer Privacy Act are categorised as data protection and privacy laws.
[89] For example, the Online Users Reduction (DETOUR) Act, the Unfair Commercial Practice Directive, Consumer Rights Directive, are categorised as consumer protection laws.
[90] Discussions predominantly surround the US Federal Trade Commission Act and its role as the regulatory body, while only a few papers mention the potential of the Digital Market Act to reign dark patterns in the EU context, and none on the TFEU.
[91] Haley (n 65).
[92] Dickinson (n 71).
[93] Lin Kyi and others, 'Investigating Deceptive Design in GDPR's Legitimate Interest', *Proceedings of the 2023 CHI Conference on Human Factors in Computing Systems* (Association for Computing Machinery 2023) <https://doi.org/10.1145/3544548.3580637>.
[94] Luguri and Strahilevitz (n 38); Calo (n 60); Hung (n 41).
[95] These differences in numbers also speak to Gray's findings that many empirical studies predominantly focus on the functionality of consent banners while subscriptions were far less commonly studied. Gray and others, 'Mapping the Landscape of Dark Patterns Scholarship' (n 8) 189.
[96] This also speaks to Susser and Grimaldi's suggestion. See Susser and Grimaldi (n 25) 250.

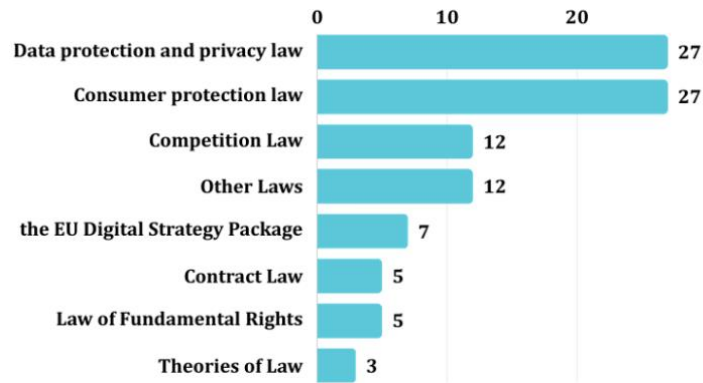

*Figure 7 Regulations mainly discussed*

### 3.3.2 What are the identified problems of the corresponding regulations?

*The fuzzy boundaries between concepts (theories of law)*

The differentiation between concepts of persuasion, manipulation, deception, coercion, and similar has been an unsettled upstream issue. It is realized that regulators must make difficult decisions regarding where the line lies precisely between these concepts in the context of dark design patterns, to eliminate problematic conduct from the market without overregulation.[97] Although dark patterns are commonly recognised as having crossed the line of proper persuasion, scholars approach the conundrum of concepts differently.

Some researchers define manipulation as hidden influences, including deception, as opposed to overt influences like persuasion and coercion.[98] Others predominantly focus on complexity between persuasion and manipulation,[99] or examine deception and dark patterns of a deceptive nature to avoid the issue of when marketing becomes unfair or abusive manipulation.[100] Some view dark patterns as coercions forcing users to spend attention, data, and money in the digital economy, exploring when persuasive and manipulative design becomes coercive in digital market competition.[101] These inconsistent approaches increase the difficulty in capturing the nature of dark patterns, which is plausibly vital for normative evaluation (See Section 4.1.1.) and results in equivocal use of terms (e.g., 'deceptive designs/patterns' or 'manipulative designs/patterns') or 'dark patterns'.

*The problem of the types and levels of 'harm' (Theories of law)*

When the conceptual boundaries between legitimate persuasion and problematic manipulation, deception, and coercion are unclear, evaluating the harm of dark patterns becomes challenging. For example, under the current US regulatory framework, many mild dark patterns (e.g., in advertisements) appear not to violate existing policies.[102] When dark patterns are subtle, traditional protective doctrines for consumers—such as duress, mistake, undue influence, misrepresentation, or *culpa in contrahendo*—are inadequate to address these issues.[103] In other words, these practices may not meet the threshold of deception as they neither make false statements nor omit material facts. They are not considered unfair as they are essentially avoidable, while the harm may be too minor to warrant limited regulatory resources.[104] The difficulty in differentiating minor harm from harms that merit legal intervention poses an obstacle for existing laws to curb harms from certain dark patterns (e.g., nagging).[105]

---

[97] Luguri and Strahilevitz (n 18) 46.
[98] Daniel Susser, Beate Roessler and Helen F Nissenbaum, 'Online Manipulation: Hidden Influences in a Digital World' [2018] SSRN Electronic Journal <https://www.ssrn.com/abstract=3306006> accessed 19 January 2024. See also Shaun B Spencer, 'THE PROBLEM OF ONLINE MANIPULATION' (2020) 2020 UNIVERSITY OF ILLINOIS LAW REVIEW.
[99] Calo (n 60).
[100] Lauren E. Willis (n 56).
[101] Day and Stemler (n 53) 34.

This lack of consensus on harm warranting intervention for dark patterns is also observed in the EU.[106] It is pointed out that promising legislations addressing dark patterns (DSA[107], GDPR[108], UCPD[109]) do not provide applicable criteria to quantify harms, nor are further guidelines issued on this matter, which hampers effective and predictable administrative decisions and individual remedies in civil proceedings. Similarly, in the EU's scenario of 'substantial injury', dark patterns in violation of EU data protection law face disparities at the national court level regarding whether remedies for 'non-material damage' should adhere to a *de minimis* damage threshold.

### *The problematic dichotomy of average and vulnerable consumers (Consumer protection law)*

Unlike the shared problems above, we identify the unique problem of each law sector hereafter. When dark patterns are backed by profiling and algorithmic personalization and go beyond uniform UI choice architecture, the outdated dichotomy of average/vulnerable consumers in consumer protection laws is identified as the crux. Building on the previous section on bounded rationality, it is increasingly recognized that the vulnerabilities of individual consumers may also be *contextual* and *influenced by external factors*, leading to the obsolescence of the traditional consumer protection rationale.[110] When dark patterns are supported by group or individual algorithmic profiling, the distinction between average and vulnerable consumers dissolves, making every consumer vulnerable as accurate detection (and even incubation) and exploitation of vulnerabilities can be easily achieved.[111]

---

[102] Eric Zeng and others, 'Polls, Clickbait, and Commemorative $2 Bills: Problematic Political Advertising on News and Media Websites around the 2020 U.S. Elections', *Proceedings of the 21st ACM Internet Measurement Conference* (ACM 2021) <https://dl.acm.org/doi/10.1145/3487552.3487850> accessed 9 December 2023.
[103] Martin Ebers', 'Liability For Artificial Intelligence And EU Consumer Law' 210.
[104] Calo (n 60) 1043.
[105] Hung (n 41) 2506–2508.
[106] Gunawan, Santos and Kamara (n 59).
[107] Regulation (EU) 2022/2065 of the European Parliament and of the Council of 19 October 2022 on a Single Market For Digital Services and amending Directive 2000/31/EC (Digital Services Act) (PE/30/2022/REV/1), OJ L 277, 27.10.2022, p. 1–102 (DSA)
[108] Regulation (EU) 2016/679 of the European Parliament and of the Council of 27 April 2016 on the protection of natural persons with regard to the processing of personal data and on the free movement of such data, and repealing Directive 95/46/EC (General Data Protection Regulation) OJ L 119, 4.5.2016, p. 1–88 (GDPR)
[109] Directive 2005/29/EC of the European Parliament and of the Council of 11 May 2005 concerning unfair business-to-consumer commercial practices in the internal market and amending Council Directive 84/450/EEC, Directives 97/7/EC, 98/27/EC and 2002/65/EC of the European Parliament and of the Council and Regulation (EC) No 2006/2004 of the European Parliament and of the Council ('Unfair Commercial Practices Directive') (Text with EEA relevance) OJ L 149, 11.6.2005 (UCPD)
[110] Fassiaux (n 31); Ebers' (n 103); Susser, Roessler and Nissenbaum, 'Online Manipulation' (n 49).
[111] Calo (n 60) 1034; Conca (n 40) 11; Faraoni (n 45).

In the US, the fall of the traditional consumer dichotomy presents significant challenges in courts. The assumptions of a 'reasonable consumer' and 'expert facial analysis' methods, such as usability tests for assessing digital interface deceptiveness, struggle with the personalized nature of algorithmic marketing, which tailors services and presentations to individuals 'in a world of one-to-one marketing', eliminating the notion of a standardised reasonable consumer. [112] Presumably the EU may face the same conundrum, as the UCPD guidance[113] has acknowledged that consumer vulnerabilities are dynamic and situational,[114] moving away from rigid categorizations based on internal characteristics. However, it's still unclear how courts will practically assess the vulnerabilities of consumers facing evolving and personalised dark patterns.

### *The problem of the consent mechanism & individual approach to data regulation (Data protection and privacy law)*

In the realm of data protection and privacy law, topics of dark patterns on consent management interfaces and their effects dominate empirical studies. The over-reliance on consent mechanisms has been criticized from many aspects. A convincing point is that, even without dark patterns, the mere repetition of dealing with different consent interfaces daily leads to consent fatigue, which undermines the informed consent mechanism intended by legislators ('fatigue by design').[115]. This consent fatigue is damaging as companies can circumvent consent with even minor dark patterns.[116] Others reiterate the indifference to the bounded rationality issue in data protection law, as policymakers fall into the 'consent fallacy', assuming people would always make decisions rationally and in their best interest with the provided information. However, people do not possess unbounded rationality, nor can they manage the full burden of information flows on data practices thoroughly, while the proliferation of dark patterns pushes the limits of the consent approach.[117]

### *The outdated theory of competition law, misposition of platform power (Competition Law)*

The problems of competition law identified are more diverse. O'Loughlin introduces the concept of 'cognitive foreclosure' and challenges the traditional rationale of competition law, highlighting how behavioural economics and the evidential effectiveness of dark patterns challenge the "strict rationality" assumption underlying competition law and question the derivative theory of efficient entry and switching related to the self-correcting capacities of markets.[118] Critics argue that outdated competition law theories result in a myopia on artificially high prices, thereby failing to promote privacy as a component of consumer welfare. This failure allows the proliferation of online manipulation by tech companies that adopt a free and low-price business model.[119] The outdated theories also lead to a mispositioning of information service platforms, creating inappropriate safe harbour exemptions for the main sources of consumer manipulation in the digital era.[120] In sum, without a theoretical review, current competition law struggles to protect consumers from dark patterns and other forms of digital manipulation.

---

[112] Lauren E. Willis (n 56) 154–157.
[113] Commission Notice – Guidance on the interpretation and application of Directive 2005/29/EC of the European Parliament and of the Council concerning unfair business-to-consumer commercial practices in the internal market C/2021/9320 OJ C 526, 29.12.2021, p. 1–129 (UCPD Guidance 2021)
[114] UCPD Guidance 2021, 100.
[115] G Spindler and L Förster, 'Privacy-Compliant Design of Cookie Banners According to the GDPR' (2023) 14 Journal of Intellectual Property, Information Technology and E-Commerce Law 2, 27; Haley (n 66) 75; Christie Dougherty, 'Every Breath You Take, Every Move You Make, Facebook's Watching You: A Behavioral Economic Analysis of the US California Consumer Privacy Act and EU ePrivacy Regulation' (2020) 12 629, 659.
[116] Hung (n 41).
[117] Sanju Ahuja and Jyoti Kumar, 'Conceptualizations of User Autonomy within the Normative Evaluation of Dark Patterns' (2022) 24 Ethics and Information Technology 52; MR Leiser, '"Dark Patterns": The Case for Regulatory Pluralism between the European Union's Consumer and Data Protection Regimes', *Research Handbook on EU Data Protection Law* (2022) <https://www.scopus.com/inward/record.uri?eid=2-s2.0-85137527810&doi=10.4337%2f9781800371682.00019&partnerID=40&md5=2f5ca8f661fa63a0201796fb0b2a2f35>; Susser, Roessler and Nissenbaum, 'Online Manipulation' (n 49).

*The new shadow cast by AI (DSA and AIA)*

Compared to aforementioned regulatory regimes, only a few papers examine potential loopholes in the EU digital strategy package, despite many highlighting its potential to regulate dark patterns.[121] A key emerging issue is whether the new laws can effectively address dark patterns influenced or driven by AI. Cohen criticizes the EU AI Act's[122] overemphasis on subliminal techniques, which are non-perceptible and highly ambiguous.[123] It argues that the real issue with manipulative AI involves attention distraction and awareness deprivation. Such abilities of AI share similarities with the essential characteristics of current dark patterns that manipulate human attentional processes, and AI can 'purposefully pull a viewer's attention toward certain visual stimuli to prevent them from attending to others,' blurring the line between unattended and imperceptible, and leading to an unreliable boundary between subliminal and non-subliminal techniques.

More importantly, this ability of AI aligns with the basic logic behind dark design patterns on websites. With the power of machine learning models, dark patterns are expected to evolve into uniform dark patterns targeting the most susceptible (vulnerable) times, or 'dynamically personalise a digital environment based on individual-level biases and vulnerabilities to inattention'.[124] Therefore, Cohen argues that the transparency requirements in the AI Act and DSA fall short since they might expose outright manipulative intent, but do not fully address manipulative stimuli.[125]

*Fundamental rights: The right to autonomy and the commercial speech*

As the harm of dark patterns to autonomy is widely recognised (See Section 3.2.4.1), some scholars argue that the absence of a fundamental right to autonomy hinders its protection against dark patterns. Both internationally and in Europe, there is a lack of consensus on which fundamental rights are violated by computational manipulation and which rights could protect individuals against such abuses, leading to calls for these rights to be explicitly defined in legislation to offer better protection against enhanced dark patterns.[126] Conversely, a regulatory challenge frequently cited in the U.S. is the First Amendment's protection of commercial speech. The proposed DETOUR Act, aimed specifically at dark patterns, and similar regulations might be obstructed by First Amendment protections, provided the commercial speech is not deemed misleading.[127] However, as discussed in earlier sections on concepts and harms, many subtle and non-deceptive dark patterns can elude the 'misleading' threshold. Consequently, companies remain advantageous in rendering the First amendment as a shield against dark pattern regulation.[128]

---

[118] O'Loughlin (n 55).
[119] Day and Stemler (n 53).
[120] Stemler, Perry and Haugh (n 32).
[121] See e.g., Conca (n 40); Faraoni (n 45); Maximilian Gartner, 'Regulatory Acknowledgment of Individual Autonomy in European Digital Legislation: From Meta-Principle to Explicit Protection in the Data Act' (2022) 8 462.
[122] European Commission, Proposal for a Regulation of the European Parliament and of the Council, 'Laying Down Harmonised Rules on Artificial Intelligence (Artificial Intelligence Act) and Amending Certain Union Legislative Acts' COM (2021) 206 final.
[123] Cohen (n 29).
[124] ibid 223–224.
[125] ibid 233.

### *The contradiction in the contracts (Contract Law)*

The efficacy of regulating dark patterns through contract law is ambiguous.[129] While some scholars argue that contracts involving dark patterns should be considered invalid because mutual assent is manipulated, the blurred boundaries between concepts still raise questions about whether contract law can guide judges in navigating these issues.[130] Moreover, the level of 'harm' still needs more empirical evidences to buttress the courts' decision against dark patterns in contracting.[131] Others critics centre the contract law doctrine, advocating for a re-evaluation of the 'freedom of contract' principle in light of how dark patterns infiltrate and secure platform-dictated standard form consumer contracts, exacerbating power imbalances and undermining the principle's original intent.[132] Furthermore, the private law nature of contract law perpetuates the ineffective individual regulatory model, in which consumers either fail to realise that they are being manipulated or tend to tolerate the cost and burden of exercising their rights through litigation, especially when class action is prima facie foreclosed by an arbitration clause.[133]

### *Limitations of the current soft law and self-regulation*

Dark pattern regulation outside of binding laws is rarely discussed, with scholars highlighting the absence of clear guidelines. Academics find it challenging to derive practical guidelines (i.e., precise specifications regarding the concrete design of digital consent tools) from complex regulations, leading to widespread legal non-compliance, such as with cookie banners,[134] and fostering environments conducive to dark patterns in privacy policies.[135] Thus, more legal intervention is deemed necessary to foster multi-dimensional solutions, as relying on industry self-regulation is impractical due to economic conflicts of interest.[136] Where soft laws are in place, their effectiveness is often questioned. Empirical studies show that self-regulatory measures like warning labels in video games are unreliable, especially on mobile platforms.[137] Consent Management Platforms (CMPs) that follow the self-regulatory transparency and Consent Framework (TCF) by IAB Europe may fail to comply with data protection laws, with some enabling the customization of dark patterns, undermining their intended purpose.[138]

---

[126] Faraoni (n 45) 11; Gartner (n 121) 472; Ahuja and Kumar (n 117). Gartner suggests that while explicitly defining a fundamental right to autonomy could enhance individual protection against dark patterns, the current phrasing in the Data Act and other European data strategy legislation already implies this right through existing human rights, such as data protection and privacy. Comparatively, Ahuja and Kumar argue that policy should differentiate between conditional and ideal autonomy, the latter being defined by law.
[127] Dougherty (n 115) 656.
[128] Luguri and Strahilevitz (n 18) 100–101; Garon (n 48) 210.
[129] Here we observed mixed opinions. For example, Dickinson (2023, p.1661) claims that private laws including tort law and contract law can remedy dark patterns sufficiently with other existing regulatory frameworks.
[130] Ebers' (n 103) 210.
[131] Luguri and Strahilevitz (n 18) 92–93.
[132] Kristin B Cornelius, 'Zombie Contracts, Dark Patterns of Design, and "Documentisation"' (2019) 8 INTERNET POLICY REVIEW.
[133] Stemler, Perry and Haugh (n 32) 650.
[134] Gerald Spindler and Lydia Fbrster', 'Privacy-Compliant Design of Cookie Banners According to the GDPR' 3.
[135] Kyi and others (n 93).

### 3.3.3 What revisions of regulations are proposed to solve dark patterns?

This sub-question focuses on concrete proposals for refining regulations or providing actionable solutions to dark patterns. Although many underlying problems lack straightforward solutions, it is noted that many proposed solutions are not tied to a specific legal domain.

#### *3.3.3.1 Proposal 1: Paradigmatic changes of legal doctrines*

Various scholars propose paradigm changes in legal doctrines to better address the challenges of dark patterns and other digital manipulations. Among these, some advocate for *lex specialis*, specifically targeting dark patterns. The proposed US Deceptive Experiences to Online Users Reduction (DETOUR) Act aims to curb manipulative dark pattern behaviour by prohibiting large online platforms from using user interfaces that intentionally impair user autonomy, decision-making, or choice.[139] The bill encourages promising regulatory initiatives, such as empowering the FTC to support a registered standard body that issues guidelines for platform design practices, oversees platforms' routine disclosure of behavioural experiments, and the establishment of independent review boards.[140] Griffin similarly proposes a specialized law targeting manipulative designs by 'systematically important platforms', introducing protective default settings, enhanced user controls through service feature disaggregation (e.g., disabling loot boxes and timed regenerations in games), and imposing extra taxes on manipulative design practices.[141]

Another approach involves updating competition law theories and follow-up policies to control platform powers. Marks introduces the concept of 'biopower' and discusses how platforms harness biopower through unregulated conglomerates, concentric mergers, and surveillance control networks to influence social norms.[142] He advocates for expanding the 'consumer welfare' concept and revitalizing merger control to collectively evaluate data flows, biopower, and coercive choice architectures, including dark patterns. O'Loughlin introduces 'cognitive foreclosure', highlighting how dark patterns by large platforms lead to market failures that justify anticompetitive intervention.[143] Day and Stemler argue that dark patterns and other online manipulations lead to anticompetitive effects akin to supra-competitive pricing, suggesting that recognizing the cost of such harm on consumer autonomy as a reflection of market quality justifies using competition law against dark patterns.[144]

---

[136] Bongard-Blanchy and others (n 6).
[137] Xiao (n 75).
[138] Habib and others (n 27).
[139] Mark Warner (n 75) 8.
[140] ibid.
[141] Griffin (n 64).
[142] Marks (n 43) 523.
[143] See n 118

Contract law and the duty of fiduciary are also viewed as promising approaches to regulate dark patterns. When traditional rules on duress and misrepresentation fall short, some researchers believe that the doctrine of undue influence in contract law offers a viable solution.[145] However, they note that the relational domination between platforms and consumers may hinder this approach, which could be addressed by recognizing 'fiduciary duties of platforms to consumers'. Supported by Hartzog & Richards, this proposal advocates for imposing 'a duty of data loyalty' on platforms, shifting the regulatory focus from proving individual harm to addressing the power imbalance and imposing more duties on platforms.[146] Similarly pinning on the imbalanced relationship, Willis advocates for reversing the burden of proof, suggesting a judicial default presumption when consumers hold false beliefs material to a transaction, assuming the business stands to benefit, and broadening the definition of 'unfair practices' to include any exploitation of consumer confusion for profit.[147]

In terms of consumer protection and data privacy laws, there is a growing demand to expand the application of the fairness principle beyond individual decision-making moments. Leiser advocates for a broader approach to tackling dark patterns through a specialized Code of Conduct, potentially extending beyond UCPD's scope, promoting a shift towards 'fairness-by-design' to hold designers accountable.[148] Additionally, linking the fairness principle from UCPD to GDPR, researchers urge lawmakers to analyse the overall system architecture, UX and UI more holistically rather than merely simplifying consent procedures.[149] Similarly, Wiener argues that privacy law should treat consent as an ongoing agency process, with assessments of unfair designs extending beyond mere agreement moments to encompass all interactions.[150]

### 3.3.3.2 Proposal 2: Refining the existing regulatory frameworks[151].

Beyond macro paradigm shifts, there are calls for refining statutes and strengthening enforcement. Given the historically weak enforcement of consumer protection law, following the proposed Dark Pattern Code of Conduct under UCPD framework,[152] Leiser advocates for establishing clearer boundaries between allowable and non-allowable commercial practices via a two-criteria assessment, and creating a blacklist for unacceptable dark patterns. In the US, efforts are focused on empowering the FTC. Wilson explores the FTC's potential to promulgate rules that explicitly define and prohibit dark patterns under Section 18 of the FTC Act.[153] He also suggests amending the Restore Online Shoppers' Confidence Act (ROSCA) to broaden its scope, thereby enhancing the FTC's ability to regulate dark patterns. Furthermore, there are suggestions that the FTC could better utilize consent decrees as a tool for ongoing supervision of companies post initial violations of Section 5, ensuring more stringent and consistent oversight of dark patterns.[154]

---

[144] See n 119
[145] Luguri and Strahilevitz (n 18).
[146] Hartzog and Richards (n 61) 1031.
[147] Lauren E. Willis (n 56) 119.
[148] Leiser (n 66).
[149] Leiser (n 117) 248.
[150] Wiener (n 81) 169.
[151] Since we do not imply design ethics as calls for self-regulation unless researchers articulate the stance to promote it as a regulation approach, we find few papers that substantially propose solely self-regulation. Rather, appeals for more support from regulators and policymakers remain mainstream. The only exception is Justin Hurwitz, 'Designing a Pattern, Darkly' (2020) 22 i.
[152] See n 148
[153] Wilson (n 26).

To bolster enforcement, many researchers argue that empirical studies and insights from HCI practitioners should play a larger role in regulatory discussions. For instance, empirical data on the effectiveness of various dark patterns could inform enforcement priorities and help define the line between permissible and impermissible practices.[155] Scholars propose that HCI researchers contribute to regulation by developing flexible taxonomies of dark patterns and automated detection methods for quicker violation checks.[156] Moreover, the effectiveness of laws and soft laws against dark patterns should be periodically reassessed through empirical legal research.[157] Meanwhile, it is proposed that regulators should consider collaboration with HCI designers, such as developing User-centred CMPs or setting up evidence-based design criteria, ensuring that only compliant CMPs enter the market.[158]

### 3.3.3.3   Proposal 3: Technical design-embedded solution and design accountability check

Possible solutions include not only amendments to legislation or new regulatory frameworks but also certain technical solutions that are positioned as a form of regulation. For example, dark pattern detection is significant contributions from HCI,[159] involving enhanced taxonomies of dark patterns used as criteria or integrated into automated detection systems. Mathur et al. suggest outsourcing the identification and flagging of dark patterns to the public, aiding the development of countermeasures like dark pattern blockers and machine-learning-powered automatic detection systems.[160]

Others address system and function redesign to combat dark patterns at their core. O'Connor et al. concentrate on redesigning the opt-out mechanisms for data protection and privacy laws, advocating for conspicuous opt-out mechanisms for personal information sales that go beyond a simple link to comply with the CCPA.[161] Comparatively, Spindler and Forster explore embedding the Do-Not-Track function within software/browsers to enhance user control over personal data sharing with online services and propose 'button labelling' solutions as counter-nudges against dark patterns.[162] They specifically propose the deployment of a 'Personal Information

---

[154] Luguri and Strahilevitz (n 18) 98–99. Other ideas on expanding the remit of the FTC are also proposed by Sylvain (2022) who calls for the FTC to take over the responsibility that is failed by the FCC and the Telecommunications Act, regarding protecting consumers from deceptive and manipulative practices by consumer-facing information services.

[155] ibid 48.

[156] Thomas Mildner and others, 'Defending Against the Dark Arts: Recognising Dark Patterns in Social Media' (Association for Computing Machinery 2023) 2372 <https://doi.org/10.1145/3563657.3595964>.

[157] n 137

[158] Midas Nouwens and others, 'Dark Patterns after the GDPR: Scraping Consent Pop-Ups and Demonstrating Their Influence', *Proceedings of the 2020 CHI Conference on Human Factors in Computing Systems* (ACM 2020) <https://dl.acm.org/doi/10.1145/3313831.3376321> accessed 19 January 2024.

[159] Papers that incorporate discussions on detection and policing are limited in number in our corpus, as they are seen as appendants to the substantial regulatory discussions which our review focuses on.

[160] Mathur and others (n 27).

[161] Sean O'Connor and others, '(Un)Clear and (In)Conspicuous: The Right to Opt-out of Sale under CCPA', *Proceedings of the 20th Workshop on Workshop on Privacy in the Electronic Society* (ACM 2021) <https://dl.acm.org/doi/10.1145/3463676.3485598> accessed 19 January 2024.

[162] Spindler and Förster (n 115).

Management System' (PIMS). This system would allow users to manage and update all privacy settings for various service providers via a central dashboard, complemented by an auxiliary data fiduciary framework. Such a system could address several persistent consent issues that facilitate dark patterns (e.g., consent fatigue, difficult-to-opt-out).

In addition to direct system redesign, some researchers explore frameworks for UX design accountability. Targeting the generally poor opt-out design for data sharing, Gunawan et al. suggest a check process on design feature parity to help enforcement agencies assess the balance of UX design 'beyond opt-in/opt-out requests, and across modalities'.[163] On the other hand, Wagner et al. believe that the German enforcement authority's 'procedural accountability' approach to Facebook's dark patterns is a promising prototype, through transparency and accountability requirements.[164] According to their construct, this procedural accountability check involves regulators requiring details on how a task is handled, which further triggers the oversight mechanism regarding the design of the business processes, workflows, and technologies by external or internal relevant parties (e.g., regulators and users).

Furthermore, when discussing 'dark patterns' in the context of rhetoric and visuals in standard-form documents, proposed solutions also involve strong technical reshaping. Cornelius advocates for the adoption of document engineering for standard form consumer contracts (including privacy and cookie policies) (SFCCs), emphasizing principles like standardization, documentation, and explanation.[165] This involves comprehensive documentation of the presentation, design, and content of documents to ensure they are preservable/scrapable, trackable/stable, authentic/reliable, and processable, supporting informed public discourse and balancing power asymmetries.[166] Similarly, Seuchter et al. call for systematic cookie-policy documentation and management, advocating for machine-readable cookie policies that can be standardized and presented consistently, with version control allowing for archived cookie policy checks.[167]

Finally, a highly popular technical solution among HCI researchers is a browser-based approach to dark patterns, especially concerning privacy and personal data protection. A browser-level consent management interface is viewed as a sustainable solution that automates user consent decisions, significantly reducing consent fatigue and effectively addressing dark patterns on cookie banners.[168] This approach could provide a scalable opt-out solution when integrated with 'Do not sell' signals, replacing inherently flawed individual opt-out methods.[169] Additionally, standardized, machine-readable cookie policies displayed via browsers would enable users to better understand and manage their consent.[170] For consumer protection beyond data privacy, browser-level solutions could incorporate community-supported automated dark pattern detection and blocking plugins.[171]

---

[163] Gunawan and others (n 24) 377:23.
[164] Wagner and others (n 27) 266.
[165] Cornelius (n 132).
[166] ibid 5.
[167] Seuchter and others (n 69).
[168] Nouwens and others (n 155) See e.g., Habib and others (n 24); Kyi and others (n 90).
[169] O'Connor and others (n 161).
[170] Seuchter and others (n 69).
[171] Mathur and others (n 27).

However, it is worth noting that while these toolkits provide valuable contributions, most are confined to privacy protection, which is a more visible phenomenon. Such technical solutions may not be generalizable to other aspects such as consumer protection. This also leads to the questions of sutibility of such solutions as to the existing legal framework.

# 4 Discussion

This section evaluates the effectiveness of existing and future regulations in solving dark pattern problems by synthesizing the most essential and constructive findings.[172] It argues that the inherent ambiguities in the concept of 'dark patterns' present significant barriers to crafting mature and effective regulations. However, certain proposed solutions offer insightful approaches for more proactive regulation. The discussion first elaborates on the intrinsic conceptual issues—the nature, harms, and scope of dark patterns—that obstruct effective regulatory measures. Then it explores how suggested technical solutions, accountability frameworks, and practical design guidelines could lead to more promising regulatory strategies. This discussion culminates in a consideration of legal pluralism as a viable approach to more effective dark-pattern regulation.

## 4.1 Uncertainties of the nature, harms and scope of the 'dark pattern' concept

### 4.1.1 The normative 'nature' problem

Effective regulation is hindered by the ambiguous nature of diverse types of dark patterns. As specified previously, the extent to which dark patterns are persuasive, manipulative, deceptive, or coercive is not well-defined. However, clarifying the normative nature of different dark patterns is crucial as they underpin the existing regulatory framework. Consumer protection law depends on interpreting deception (e.g., 'misleading commercial practices' in UCPD[173] and 'deceptive acts or practices' in FTC Act[174]) and, to some extent, coercion ('aggressive commercial practices' in UCPD[175] and 'unfair acts or practices' in FTC Act[176]). Moreover, interpreting coercion is essential to justify interventions from competition law. Unfortunately, despite frequent references to 'manipulative' or 'deceptive designs' as synonyms for dark patterns, explicit concept differentiation is lacking in existing laws addressing dark patterns, creating a barrier to regulation.

Several factors contribute to this normative ambiguity. Firstly, even theoretically, the differentiation of these concepts is unresolved. Secondly, dark patterns, by nature, are a collective and thematic concept;[177] some types may be deceptive (e.g., sneaking into baskets, disguised ads) while others are coercive (e.g., nagging). More importantly, considering the rapid evolution and diversification of dark patterns, case-by-case determination of their normative natures, especially in *ex-post* paradigms, could increase the regulatory resource burden. Thirdly, the accumulation of various dark patterns throughout UX design could complicate distinguishing their nature. When reviewed against the backdrop of more advanced dark pattern forms, discernibility could be further obscured.

---

[172] As the inherent problems of each law (Section 3.3.2) and worth their own research, and many root problems (Section 3.2.3) may incur bigger discussion beyond regulation, we only discuss them in this section when necessary.
[173] Articles 6 and 7, UCPD.
[174] Section 5, Federal Trade Commission Act (FTC Act) 1914 s 5
[175] Article 8 and 9, UCPD
[176] n 174
[177] Mathur, Kshirsagar and Mayer (n 4).

Consequently, the unsettled nature of dark patterns complicates regulation. Further research could enrich the normative discourse by, for example, discussing the difference between 'coercion' as a simple design feature (e.g., nagging) and as a contextual barrier to switching services (lock-in effect), or even as the absence of better-designed alternatives (change of social norm), and explore how laws (e.g., consumer protection, competition, human rights) might interplay. Nonetheless, it might be more pertinent to question whether rigidly categorizing design patterns into traditional legislative structures could achieve effective regulatory outcomes, especially when these legal frameworks may themselves require re-evaluation in the digital era (See 3.3.2). Therefore, dark pattern regulations that rigidly assign fixed normative 'labels' to each identifiable type may not be practical. The regulatory discussion could benefit from exploring a more holistic concept of 'unfairness' and 'unfair patterns' (potentially with more ex-ante paradigms), which could encompass all unfair influences without getting stalled at the preliminary phase of concept subdivision.[178]

### 4.1.2 The 'harms' problem

Related to the problem of concept differentiation, current dark pattern regulations struggle due to the difficulty in identifying evident and actionable harm. The blurred line between acceptable persuasion and unacceptable dark patterns complicates the classification of harm magnitudes and clarification of which levels merit legal remedies. While harm types may seem clear, it is important to note that, aside from financial harm, other types are intangible harms that are difficult to quantify and demonstrate. This is highlighted by Gunawan et al.'s findings, where the absence of damage claims in GDPR case law is concerning given the potential severity of dark pattern harms.[179] Although financial harm is theoretically easier to capture, the lack of proactive enforcement of consumer protection at the EU member state level and the scarcity of case laws against mild dark patterns (e.g., hidden costs) provide insufficient references.[180] Thus, basing dark pattern regulation on the demonstration of quantifiable and evident harm may hinder effective regulation.

*The discrepant theories on harm to the market and competition*
The varying interpretations of harm to market and competition limit regulatory approaches to dark patterns. Although literature generally acknowledges dark patterns' harm to the market and competition by imposing anti-competitive effects, diminishing consumer welfare, and eroding market trust (See Section 3.2.4.2), opinions differ on how anti-competition effects are incurred and how harm to consumer welfare interacts in this process.

---

[178] This would raise questions about the concept of 'unfairness' (similarly, 'fairness'), which needs to be evaluated against the culture and expectation within the community that is influenced by the most pertinent dark pattern forms. In this sense, sector-based empirical evaluation of '(un)fair' designs could be indispensable.
[179] Gunawan, Santos and Kamara (n 59).
[180] Leiser (n 66) 513.

Some scholars emphasize the role of platforms with market power, arguing that their dominance transforms mild design patterns from mere persuasion to coercion and market abuse. This manipulation steers users towards suboptimal behaviours, creating cognitive foreclosure and consumer-side market failure that damage consumer welfare and market competition.[181] Conversely, other researchers argue that market power might be irrelevant, and that the use of dark patterns may not significantly impact market competition. They suggest that dark patterns, although subtle, effectively steer consumer decisions without necessarily causing backlash.[182] Consumer biases are so readily exploited that firms typically focus on capitalizing on these biases and vulnerabilities rather than competing to 'debias consumers', leading to a 'race to the bottom' in market competition.[183] Consequently, although economic harm at the market level may be ambiguous, the mechanism of harm at the consumer level is clear and likely to persist over time.[184]

However, these two perspectives may already be occurring concurrently, collectively reducing overall consumer welfare, regardless of the contested anti-competitive effect. Under traditional competition law theories, if consumer protection is considered secondary to market competition protection, the remedial potential of competition law against dark patterns remains limited. An interesting area for further research is the implication of the EU DMA,[185] which does not directly address the causality of protected value but notably imposes an ex-ante ban on interface designs by gatekeepers that could undermine the effectiveness of prohibitions and obligations specified in the Act,[186] which also speaks to our findings on the harm to the functions of law in Section 3.2.4.3. Therefore, DMA seem to acknowledge that steering users' decision-making by entities with great market power ('gatekeepers') through dark patterns would result in contestability reduction in the digital market, which in itself is worth regulating.

### 4.1.3 The 'scope' problem

The effectiveness of regulation is impacted by the expanding scope of the dark pattern concept in legal scholarship (See 3.2.2). This expansion could be significant but also poses risks that might affect regulatory effectiveness for two reasons.

Firstly, an expanding scope of dark patterns could dilute its distinctiveness, potentially merging the term with the broader context of online/digital manipulation. Originally, 'dark patterns' referred primarily to surface UI manipulations; The strong initial focus might render broader interpretations of the term somewhat overstretched. Therefore, Sánchez Chamorro et al. criticize the amalgamation of definitions and terms used to describe 'dark patterns' and suggest shifting the discussion to a more precise term such as 'manipulative designs'.[187] This term differentiation aids in shifting the focus from specific patterns to broader concerns about entire systems, design strategies, and contextual manipulation. Retaining the original understanding of dark patterns as design tricks used on apps and websites at the surface UI level, while using other terms for systemic and functional, personalized designs, could prevent term dilution and confusion, ultimately supporting more effective regulation. However, achieving clarity is unlikely to be easy, given the continuously diversified interaction formats (e.g., via VAs, VR) between users and digital services, which extend beyond traditional graphical UIs to a systemic level in practice. Additionally, the broad interpretation of 'interface' as 'any software' in DSA appears to anticipate such conceptual dilution.[188] Nevertheless, regardless of the uncertainty around terminology which necessitate more future legal clarification, the term 'dark patterns' for now should be applauded and valued for its role in raising public awareness about the risks of unchecked imbalanced design power wielded by digital choice architects.

---

[181] Day and Stemler (n 53). O'Loughlin (n 55).
[182] Luguri and Strahilevitz (n 18).
[183] Barros and others (n 47) 141.
[184] Calo (n 60) 1030.
[185] Regulation (EU) 2022/1925 of the European Parliament and of the Council of 14 September 2022 on contestable and fair markets in the digital sector and amending Directives (EU) 2019/1937 and (EU) 2020/1828 (Digital Markets Act) (Text with EEA relevance) (PE/17/2022/REV/1), OJ L 265, 12.10.2022, p. 1–66 (DMA)
[186] Recital 37, 70. Article 13 (4) and (6).

Secondly, an expanding scope might also complicate regulatory decision-making concerning the ability to retrospectively classify some existing system-level designs as dark patterns, as these practices may have become normalized. For instance, HCI researchers might view targeted advertising or algorithmic/discriminative pricing as analogous to dark design patterns and express concerns that such designs could lead to excessive 'wealth surplus extraction' detrimental to consumer welfare, yet this type of harm is not prominently addressed in legal discussions.[189] Further research might explore other dark pattern types that reflect inherent discrepancies in value assessment between legal and HCI researchers.

In summary, the ongoing expansion of the dark pattern scope could hinder effective regulation. Given the interdisciplinary nature of 'dark patterns', a review of how the term is used and a strategic approach to addressing the boundary issue would be best determined collaboratively by HCI and legal researchers.[190]

## 4.2 Promising technical and regulatory solutions

### 4.2.1 Compatible technical solutions and design accountability check frameworks

Despite the inherent conceptual problems of dark patterns that make current regulations appear ineffective, our findings on proposed solutions highlight promising routes to more effective future regulation. These solutions take the form of technical system/function redesigns or supervision and standardization of online choice architecture that are proactive and reflective of legal principles such as 'privacy by design and by default' and 'fairness'. (See Section 3.3.3.3).

---

[187] Lorena Sánchez Chamorro, Kerstin Bongard-Blanchy and Vincent Koenig, 'Ethical Tensions in UX Design Practice: Exploring the Fine Line Between Persuasion and Manipulation in Online Interfaces' (Association for Computing Machinery 2023) 2418 <https://doi.org/10.1145/3563657.3596013>.

[188] Weiwei Yi, 'Gaming the Mind: Unmasking "dark Patterns" in Video Games' [2024] Internet Policy Review <https://policyreview.info/articles/news/unmasking-dark-patterns-video-games/1739> accessed 2 July 2024.

[189] See e.g., Narayanan and others (n 25); Susser and Grimaldi (n 25).One possible explanation could be that practices such as differentiated pricing can be seen as pro-competition and fall into the scope which traders have the right to decide. See Organisation for Economic Co-operation and Development, 'Personalised Pricing in the Digital Era' (OECD, DAF/COMP (2018)13, 20 November 2018) < https://one.oecd.org/document/DAF/COMP(2018)13/en/pdf >. Therefore, the practice of algorithmic pricing and the extended extraction of wealth surplus per se is unlikely a 'legally regulatable' dark design pattern.

[190] There have been voices from academics calling for and contributing to a unified ontology of dark patterns, which could enhance the regulability of dark patterns, especially for types that have been widely recognized and share many common characteristics. This could be a promising strategy to address the most salient dark patterns first before regulating the deeper forms. See CM Gray, C Santos and N Bielova, 'Towards a Preliminary Ontology of Dark Patterns Knowledge' (2023) <https://www.scopus.com/inward/record.uri?eid=2-s2.0-85158166826&doi=10.1145%2f3544549.3585676&partnerID=40&md5=c99a14db735729fc1cca235b2c53a2a5>.

Direct technical design-embedded solutions (e.g., Personal Information Management Systems (PIMs), document engineering, browser-based approaches) collectively envision a standardized and centralized decision-making environment. This could be achieved through a hub-like central dashboard or more feasibly through a browser-based consent management interface. Despite longstanding critiques of the current data protection and privacy laws' focus on individual consent, the real issue is the decentralized, unstable, one-directional (i.e., hard to opt-out) decision-making architectures that lead to consent fatigue, thus fostering and amplifying the effectiveness of dark patterns and, consequently, the failure of the consent mechanism. In other words, it is the lack of a stable, standardized, centralized, reviewable, user-friendly decision-making environment that hampers the efficacy of data protection and privacy laws and the consent mechanism. Matched supportive document engineering enables the systematic tracking, analysis and comparison of SFCCs and the affiliated dark patterns of and between the companies' further support more informed consent. Despite the promise of these technical solutions, they predominantly focus on privacy-related issues, which is only one aspect of dark patterns. This narrow scope overlooks other significant legal values and scenarios affected by dark patterns, such as consumer protection and competition. Consequently, while privacy protection may be enhanced, other forms of manipulative design practices may remain unaddressed.

On the other hand, indirect accountability assessment frameworks for UX design (e.g., the 'feature parity check' for opt-out mechanisms, 'procedural accountability' assessment) align well with the paradigmatic changes in consumer protection and privacy law doctrines (See Section 3.3.3.1), where legal researchers call for expanding the assessment scope to UX design beyond the consent moment to achieve fairness by design.[191] This concurrence suggests that a broader assessment framework for UX design against dark patterns is gaining theoretical momentum.

While the browser-based approach is promising and popular, conceptualized in the proposed e-Privacy Regulation (Articles 9 and 10),[192] more research is needed on the actual subjective that undertakes the responsibility to reshape the infrastructure of data management. For instance, the term 'browser' implies a burden on browsers (as the portal to websites) but overlooks their counterparts—apps—on computer device operating systems. Further research should discuss who should bear the responsibility (apps, OS, or browsers), in addition to the justification for imposing responsibilities on them.

---

[191] Leiser (n 66); Wiener (n 81).
[192] European Commission, Proposal for a Regulation on Privacy and Electronic Communications (10 January 2017) <https://digital-strategy.ec.europa.eu/en/library/proposal-regulation-privacy-and-electronic-communications> accessed 30 May 2024

Although both laws aim to tackle similar problems related to individual decision-making susceptible to manipulative choice architectures, there is a notable disparity in the quantity of technical solutions addressing data protection and privacy versus consumer protection.[193] For example, only Mathur et al. discuss using publicly outsourced flagging and machine-learning-based classifiers in browsers to curb dark patterns on shopping websites as a regulatory approach substantially among the papers on consumer welfare we reviewed.[194] This disparity in quantity suggests that the protective paradigm of consumer protection law might inherently be weaker than that of data protection and privacy. The 'privacy by design and by default' principle justifies stronger measures to guarantee the default status of privacy, endorsing more enterprising technical solutions to discipline the choice architectures regarding personal data and privacy. Comparatively, consumer protection law relies upon the understanding of (transactional) fairness between traders and consumers, where users/consumers are more likely to be left in the wild, as the line deemed fair as well as the assessment scope is not uniform for individuals, while the universal lack of reflective investigation on consumers' attitudes to dark patterns in scholarship in this sector holds back the effective regulation of dark patterns through consumer protection law. Consumers are left to rely on self-regulation and self-help remedies (as suggested by Mathur et al.), but these are likely blocked by the lack of a 'right to repair' that could break through IP protections on software/interface design.[195] Ultimately, even for self-help remedies, people lack the tools and legal support to defend themselves against dark patterns and the inherently unilateral power of design.

### 4.2.2  Practical design guidelines cross regimes

The criticisms regarding the sufficiency of existing regulations converge on the uncertainty of how to determine whether a design employs dark patterns, leading to a common call for clear guidelines/guidance on assessment or design.[196] For instance, industry-led ethical design guidelines of a self-regulation nature could serve as a baseline in courts to determine whether 'professional diligence' is fulfilled, according to which a design can be judged as 'unfair' ('dark') or not under UCPD.[197] Similarly, such guidelines can take other forms (e.g., Code of Conduct, Guidance, Recommendations) issued directly by authorities.[198] Fortunately, this is an ongoing movement in the EU (e.g., EDPB Guideline 2022)[199] and we expect to see more according to Art 25(3) DSA. The only caveat is that regulators need to evaluate and scrutinize the quality and compliance of industry-led guidance (self-regulation), considering the inherent interest conflict.[200]

It is worth highlighting that the line between what is appropriate design versus dark design is not fixed and permanent. According to Bernstein's theory on the window of reshaping technology, the status of no regulation continually changes the norm in the commercial environment day by day, where consumers might become more tolerant of easy-to-perceive and popular dark patterns, making them a new normal.[201] However, this assumption applies only to "average" consumers; for "vulnerable" consumers who are less adaptive to the change of norm, the basic types could still be exploitative. This also explains why, especially in the context of consumer protection law, taking the initiative to form rules and 'cut the line' might be more effective than 'finding the line'. The more hesitant policymakers are, the more likely the overall line will gradually shift to a side where more consumer welfare is compromised (especially in the form of unmeasurable harm), especially considering the already imbalanced relationship between consumers and companies that deploy dark patterns.

---

[193] This discrepancy also speaks to the previous findings that consumer protection law is more criticised for the interpretation of provisions, while data protection and privacy law is more for their enforcement quality.
[194] Mathur and others (n 27).
[195] Kollnig and others (n 70).
[196] As the regulations to dark patterns seem to scatter in many different laws (as our finding shows), the question of how to reconcile these laws and arrange the hierarchy of legal forces, ex-ante or ex-post is outside the scope of our research.
[197] Conca (n 40). and UCPD Guidance 2021, 101.
[198] Sánchez Chamorro, Bongard-Blanchy and Koenig (n 183) See e.g., Leiser (n 63).
[199] European Data Protection Board, Guidelines 3/2022 -Version 2.0 on Dark patterns in social media platform interfaces: How to recognize and avoid them, 24 February 2023
[200] Xiao (n 75).

## 4.3  Legal pluralism as a potential direction

To develop more effective regulation against the backdrop of uncertainties surrounding dark patterns, it is essential to adopt a comprehensive regulatory paradigm. Essentially, dark patterns represent the under-scrutinized unilateral power of design in digital choice architecture, which has been widely recognized at the UI level, compared to the emerging use of the term in a more systemic level and broader context. Even without considering the expansion of the connotation of dark patterns and the inherent uncertainty of the term 'dark patterns', the findings section of the paper already demonstrates that the harm caused by dark patterns impacts various legal interests and values addressed by different regulatory mechanisms. However, the current regulatory system remains limited both theoretically and in terms of its outdated approach to tasks. For example, it is challenging to differentiate between concepts, including persuasion and manipulation, and to evaluate the types and levels of harm caused by dark patterns. Meanwhile, outdated consumer protection frameworks that are not designed for algorithmic personalization, over-reliance on consent mechanisms in data protection and privacy laws, and inadequate competition laws that fail to address digital manipulations contribute to the ineffectiveness of the current regulatory landscape.

While both HCI and legal scholarship endeavour to grapple with these issues, a more comprehensive framework is required to take advantage of both strengths. In this sense, the power of design highlighted by Woodrow Harzog[202] further speaks to Lessig's theory of 'Code is law',[203] which calls for recognition of the fact that technologies also reflect value and calls for a shift in societal mindset to recognize the need for collaboration between technology and regulations to ensure commonly recognized values in cyberspace. Such legal pluralism can effectively bridge the gap between technological advancements and regulatory frameworks, ensuring that the harmful impacts of dark patterns are mitigated. Accordingly, regulating dark patterns would not only require collaboration between different legislative frameworks, regardless of whether it's data protection and privacy law, consumer protection law, or even competition law, but also necessitate focusing on core legal objectives. These objectives are not merely to distinguish between 'data subjects' and 'consumers' but to harness the strengths of one regulatory mechanism to ensure fairness and autonomy while compensating for its weaknesses with others. This is exemplified by the recent EU digital regulatory package, such as DSA, DMA, and AI Act.

---

[201] Bernstein (n 1).
[202] Woodrow Hartzog, *Privacy's Blueprint: The Battle to Control the Design of New Technologies* (Harvard University Press 2018).
[203] Lawrence Lessig, 'Code Is Law' (2000) 1 Harvard magazine 2000.

Additionally, it is essential to consider interdisciplinary approaches that can achieve legal objectives or be integrated into legal governance frameworks. For example, the taxonomy of dark patterns and related automated detection methods developed by HCI scholars can be utilized by regulators.[204] These methods can also be embedded into system design to promote fairness by design and privacy by design. Furthermore, as previously mentioned, soft laws, such as designer guidelines and codes of conduct, should also play a regulatory role. On the one hand, such measures can preserve the necessary autonomy of designers for various services in practice. On the other hand, they can provide flexible and adaptable frameworks that encourage best practices and ethical standards, thereby fostering a culture of responsibility and accountability within the industry. Besides using soft measures, in some areas such as the use of personal data, a more stringent approach should be adopted to ensure a standardized, stable, and reviewable environment through centralized personal data management panels, with the support of document engineering of disclosed information.[205] Only through such legal pluralism can dark patterns be more effectively regulated.

# 5 Limitations

Despite the review choosing the most pertinent keywords of dark patterns and disciplines for database searching, some pertinent papers might not be enlisted due to the different use of terms or indexing of disciplines. Additionally, as we only focus on English-written scholarship, this article might not reflect dark patterns of research written in another different linguistic background. Meanwhile, the difficulty in tagging some papers, especially those HCI researcher-led papers to which law researchers substantially contributed, as either law or HCI due to the ongoing deepening collaboration in dark pattern research from an interdisciplinary perspective, and consequently, our strategic inclusion of these papers as law papers to enrich our analysis may, from another perspective, leading to less heterogenous findings for RQ2. We did not systematically identify the actors (e.g., platforms, designers, legislators and enforcers) involved in dark pattern regulation and the proposed responsibilities upon them, which may be suitable for future research on a more micro and practical level.

# 6 Conclusion

In this paper, we systematically reviewed research on dark pattern regulation across HCI and legal fields using mixed analysis methods. Our research highlights the diverse framing contributions and contextual themes within the literature on dark pattern regulations across disciplines. We found that legal papers heavily rely on definitions from HCI scholarship and recently attempt to adopt an expansive interpretation of the dark pattern concept to systemic issues. This paper delineates legal scholars' views on the root problems of dark pattern practices and demonstrates how dark patterns are entrenched in the digital environment, extending beyond mere online replications of offline unfair practices.

---

[204] Mathur and others (n 27).
[205] Cornelius (n 132).

We identified various harm types of dark patterns from individual, market, and social perspectives. Although privacy and data protection law and consumer protection law are the primary sectors for regulatory discussion, competition law, contract law, and many other legal frameworks are also seen as tools against dark patterns in different contexts. Despite the clarity of harm types and the variety of laws involved, universal problems across sectoral laws—such as theories, interpretation, and enforcement—and unique problems within each sectoral law pose significant challenges to effective dark pattern regulation.

Nevertheless, we synthesise three groups of proposed regulatory solutions for dark patterns. The first group includes paradigmatic changes in legal doctrines such as *lex specialis*, targeting platform power through updated competition law rationale, restoring a balanced user-platform relationship through contract and duty of fiduciary, and fairness by design through extended consumer and privacy protection laws. The second group involves refining the existing regulatory framework by empowering enforcement authorities through law amendments, complemented by empirical studies and collaborations with HCI designers. The third group focuses on technical design-embedded solutions, including automated detection and policing, privacy setting redesign, UX design accountability checks, and document engineering.

Ultimately, we argue that the inherent ambiguities in the concept of 'dark patterns'—the nature, harms, and scope—present significant barriers to crafting mature and effective regulations. We suggest that in the context of evolving dark pattern formats, regulatory discussion could benefit from exploring a more holistic concept of 'unfairness' and 'unfair patterns' to avoid getting stalled at the phase of concept subdivision of persuasion, manipulation, deception, and coercion. Moreover, more work needs to be done to conquer the over-reliance on quantifiable and evident harm in convicting dark patterns in the current regulatory paradigm, while dark patterns represent a good opportunity for reflections on competition law theories. We also argue that the ongoing expansion of the dark pattern scope in legal scholarship could hinder effective regulation and that the strategic use of terms would benefit from a collaborative review by HCI and legal researchers, to prevent concept dilution but also cover the advanced dark designs of system level and intertwined with new factors such as AI. Eventually, we highlight certain proposed solutions, such as technical design-embedded solutions, accountability frameworks, and practical design guidelines, and argue that they offer promising routes for more proactive regulation of dark patterns. We suggest that scholars and regulators consider legal pluralism as a viable approach to more effective dark pattern regulation, promoting 'privacy by design' and 'fairness by design' and beyond through better collaboration between laws, laws and technical designs, HCI designers and regulators.

# 7 Appendix

*Codebook for content analysis:*

| Codebook for (Content Analysis) | |
|---|---|
| **HCI papers** | **Law papers** |
| **RQ1** | |
| *Research frames and aims (Section 3.1.1)* | |
| a. Evaluative: leveraging existing taxonomies to identify whether something is an example of a dark pattern | I. Conceptualisation of concepts: the act could be (re)conceptualisation, re- and deconstruction of a concept. It could be the concept of "dark pattern" per se, or concepts that are related to dark patterns. |
| b. Descriptive: using examples to illustrate power, impact, or attributes | II. Assessing the existing law: assessing the current legal framework, or the lack of regulation regarding a phenomenon. The conclusion could be negative, positive, or mixed (e.g., effectiveness). |
| c. Detection-Focused: creating and deploying an automated detection technique | III. Assessing the legitimacy of a practice: could be practices or an overall phenomenon that embraces or is a component of dark patterns. |
| d. Taxonomy-Building: defining new types of dark patterns or consolidating existing patterns | IV. Proposing problem-solving schemes: the problem solving could be technical, legal, or paradigmatic, and so on, and the objective could be law per se, or related to dark patterns. |
| e. Problematising: identifying limitations of taxonomies, identifying gaps in current literature | |
| f. Experimental/Causal: identifying generalisable causal mechanisms | |
| g. Non-empirical/Theoretical: papers that do not have empirical elements | |
| *Jurisdictions (Section 3.1.2)* | |
| a. United States | |
| b. European Union (including member states) | |
| d. Multi-jurisdiction | |
| e. United Kingdom | |
| f. Canada | |
| g. China | |
| *Contextual Themes (Section 3.1.3)* | |
| a. Platform effects: When the paper particularly discusses the roles and effects of platforms or similar subjects in dark pattern-related practices. | |
| b. Digital Services: When the context is general and does not fall into other codes, or when a macro perspective is taken. | |
| c. Games and gaming: When mainly focusing on players, game designs, and designers. | |
| d. Websites and apps: When the focus is particularly detailed on user interaction with user interfaces at a micro level (e.g., Cookie banners, the check-out process on shopping websites). | |
| e. Artificial Intelligence: When the factor of AI is a major discussion in the paper. | |
| **RQ2** | |
| N/A | *How is 'dark patterns' engaged in the literature (Section 3.2.1)* |
| | a. As an example: The mention or discussion on dark patterns is limited, usually appearing as scattered sentences or paragraphs. |
| | b. As a case/theme: Dark patterns are analyzed relatively independently, as one of the cases/themes in the paper. |
| | c. As the main focus: Dark patterns are the main focus of the research (e.g., as an object of regulation, assessment of taxonomies). |
| | d. As the context/backdrop: Dark pattern discussion is set up as the background that ushers in the discussion of other focuses. |
| | *Engagement with definitions and taxonomies (Section 3.2.2)* |
| | a. Does not engage with taxonomies |
| | b. As examples of the dark pattern phenomenon |
| | c. As the objective of analysis |
| **RQ3** | |
| *The main regulations regarding dark patterns discussed (Section 3.3.1)* | |
| Data protection and privacy law | |
| Consumer protection law | |
| Competition law | |
| Contract law | |
| The EU Digital Strategy package | |
| Other laws | |
| Law of Fundamental rights | |
| Theories of Law | |